\documentclass[final,3p,times,twocolumn]{elsarticle}
\pdfoutput=1
\usepackage{hyperref}

\usepackage{graphicx}
\usepackage{upgreek}
\usepackage{physics}
\usepackage{amsmath}

\usepackage{subcaption}
\usepackage[load-configurations=abbreviations, load=addn]{siunitx}
\DeclareSIUnit{\molecules}{molecules}
\DeclareSIUnit{\TT}{T_{2}}

\usepackage{amssymb}

\usepackage{lineno}

\graphicspath{{figures/}}

\journal{Vacuum}

\begin{document}
	
	\begin{frontmatter}
		
		
		
		\title{Time-dependent simulation of the flow reduction of D$_2$ and T$_2$ in the KATRIN experiment}
		
		
		\cortext[cor1]{Corresponding authors}
		\author[adrlabel4]{F. Friedel\corref{cor1}}
		\ead{fabian.friedel@kit.edu}
		\author[adrlabel4]{C. R\"ottele\corref{cor1}}
		\ead{carsten.roettele@kit.edu}
		\author[adrlabel4]{L. Schimpf\corref{cor1}}
		\ead{lutz.schimpf@kit.edu}
		\author[adrlabel4]{J. Wolf}
		\author[adrlabel4]{G. Drexlin}
		\author[adrlabel4]{M. Hackenjos}
		\author[adrlabel3]{A. Jansen}
		\author[adrlabel3]{M. Steidl}
		\author[adrlabel3]{K. Valerius}

		\address[adrlabel4]{Institute of Experimental Particle Physics (ETP), Karlsruhe Institute of Technology (KIT), Wolfgang-Gaede-Str. 1, 76131 Karlsruhe, Germany}
		\address[adrlabel3]{Institute for Nuclear Physics (IKP), Karlsruhe Institute of Technology (KIT), Hermann-von-Helmholtz-Platz 1, 76344 Eggenstein-Leopoldshafen, Germany}

		\begin{abstract}
			The KArlsruhe TRItium Neutrino experiment (KATRIN) aims to measure the effective electron anti-neutrino mass with an unprecedented sensitivity of 0.2\,eV/c$^2$, using $\upbeta$-electrons from tritium decay. Superconducting magnets will guide the electrons through a vacuum beam-line from the windowless gaseous tritium source through differential and cryogenic pumping sections to a high resolution spectrometer. At the same time tritium gas has to be prevented from entering the spectrometer. Therefore, the pumping sections have to reduce the tritium flow by at least 14 orders of magnitude. This paper describes various simulation methods in the molecular flow regime used to determine the expected gas flow reduction in the pumping sections for deuterium (commissioning runs) and for radioactive tritium. Simulations with MolFlow+ and with 
an analytical model are compared with each other, and with the stringent requirements of the KATRIN experiment.
		\end{abstract}

		
	\end{frontmatter}
	
	\section{Introduction}

The {\bf KA}rlsruhe {\bf TRI}tium {\bf N}eutrino experiment (KATRIN) has been
designed to determine the effective mass of electron anti-neutrinos with an
unprecedented sensitivity of \SI{200}{\milli \electronvolt \per c^2} at 90\% CL, using electrons from 
tritium $\upbeta$-decay~\cite{KATRIN2005,Drexlin2013}. The analysis is focused on
the last few eV below the 18.6\,keV endpoint of the $\upbeta$-spectrum.

\begin{figure*}
	\centering
	\vspace{10pt}
        \includegraphics[width=\textwidth]{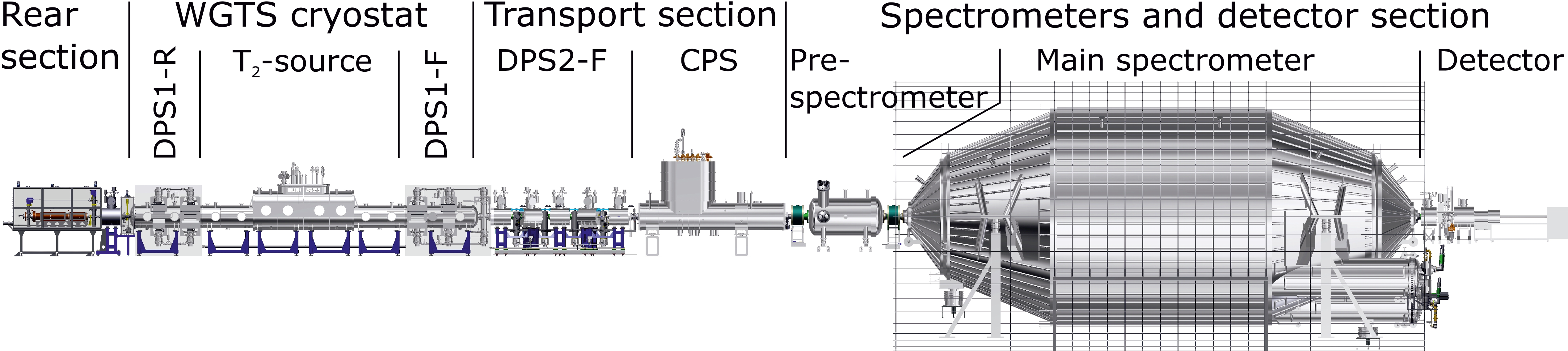}
        \caption{Overview of the 70-m long KATRIN~experiment. Tritium gas is injected in the source (WGTS) 
		and pumped out in adjacent pumping sections (DPS1/2, CPS). Electrons from $\upbeta$-decay 
		are magnetically guided to the electrostatic spectrometers, analyzing their energy, and 
		are counted at the detector.}
        \label{fig:katrinBeamline}
\end{figure*}

The experiment is located at the Karlsruhe Institute of Technology (KIT), 
Campus North, near Karlsruhe, Germany.
The approximately 70-m long beamline is shown in Fig.~\ref{fig:katrinBeamline}.
The experiment can be divided into two main sections.
The Source and Transport Section is responsible for the production and adiabatic 
transport of tritium $\upbeta$-particles to the Spectrometer and Detector Section. 
Their energy is determined in the integrating, electrostatic Main Spectrometer, 
which can provide high energy resolution with a  
wide open solid angle acceptance for $\upbeta$-electrons, emitted isotropically in the 
windowless gaseous tritium source (WGTS). The expected signal rate in the last eV of 
the $\upbeta$-spectrum is about $10^{-2}$~counts per second; thus, a necessary precondition of 
reaching the neutrino mass sensitivity goal is a similarly low background rate~\cite{KATRIN2005}.
This stringent requirement entails a thorough understanding and mitigation of background sources 
along the entire beamline of the experiment. Of particular importance are background 
electrons produced in the main spectrometer. One such background source are $\upbeta$-decays of 
tritium passing from the WGTS through the connecting beam-line into the spectrometer.  

The WGTS has been designed to produce more than $10^{11}$~$\upbeta$-particles 
per second~\cite{Babutzka2012}. A constant flow of 95\% pure T$_2$ gas is injected at the center 
of the 10-m long beam-tube with a pressure of $3.4 \times 10^{-3}$\,mbar at a temperature 
of 30\,K~\cite{arXivKuckert2018}. Superconducting magnets adiabatically guide half the emitted 
electrons through the differential~\cite{PhDKosmider2012} and cryogenic~\cite{Gil2010, PhDJansen2015} 
pumping sections towards the spectrometer section, 
while reducing the tritium flow by at least 14~orders of magnitude~\cite{KATRIN2005}. 

Over 99\% of the gas is already pumped out by the  
turbo-molecular pumps (TMP) of the first stages of the differential pumping systems, 
which are integrated at both ends of the source cryostat (DPS1-R and DPS1\nobreakdash-F). The detailed 
rarefied gas flow simulation has been described by Kuckert et al.~\cite{arXivKuckert2018}.
In the remaining text, WGTS refers to the entire source cryostat, including the 
DPS1-R and DPS1-F subsections. 
In this paper, we describe the simulation in the molecular flow regime of the flow 
reduction factors for tritium in the second stage of the differential pumping section (DPS2-F), 
and in the cryogenic pumping section (CPS).  

Section~\ref{sec:pumping_section} describes the general design and the vacuum 
system of both pumping systems. The Test Particle Monte Carlo (TPMC) 
simulation with MolFlow+~\cite{lit:molflow} is described in Sec.~\ref{sec:TPMC_models}. 
The large flow reduction along the beam tube forced us to subdivide the simulation of a long 
beam-tube into several independent steps and concatenate the results in the subsequent 
analysis. In the analysis of the CPS simulation the time dependence of the reduction 
factor was introduced, assuming a slow migration of the adsorbed and redesorbed tritium 
towards the spectrometer section. An alternative simulation method for the CPS is introduced in 
Sec.~\ref{sec:analytic_model}. The geometry is hard-coded in C++, optimizing the speed of the 
simulation. It also allows the simulation of time-dependent properties that are difficult  
to characterize with MolFlow+ in a single pass. In Sec.~\ref{sec:results}, 
the results are presented and discussed in Sec.~\ref{sec:discussion}.

	\section{The Pumping Sections \label{sec:pumping_section}}
\subsection{The Differential Pumping Section}

The Differential Pumping Section (DPS2-F) has to fulfill three different tasks. The first task is 
to reduce the tritium gas flow between the WGTS and the CPS by 5 orders of magnitude. The second 
task is to guide the $\upbeta$\nobreakdash-electrons adiabatically from the WGTS downstream to the Cryogenic 
Pumping Section (CPS) along the magnetic flux tube. Each tritium decay in the WGTS ionizes on 
average about 10 tritium molecules, which are also guided by the magnetic flux tube through the 
beamline. The third task is to prevent these ions from reaching the spectrometer section, where 
they would increase the background rate.

\subsubsection{Geometry}

\begin{figure}
	\begin{center}
		\includegraphics[width=\linewidth]{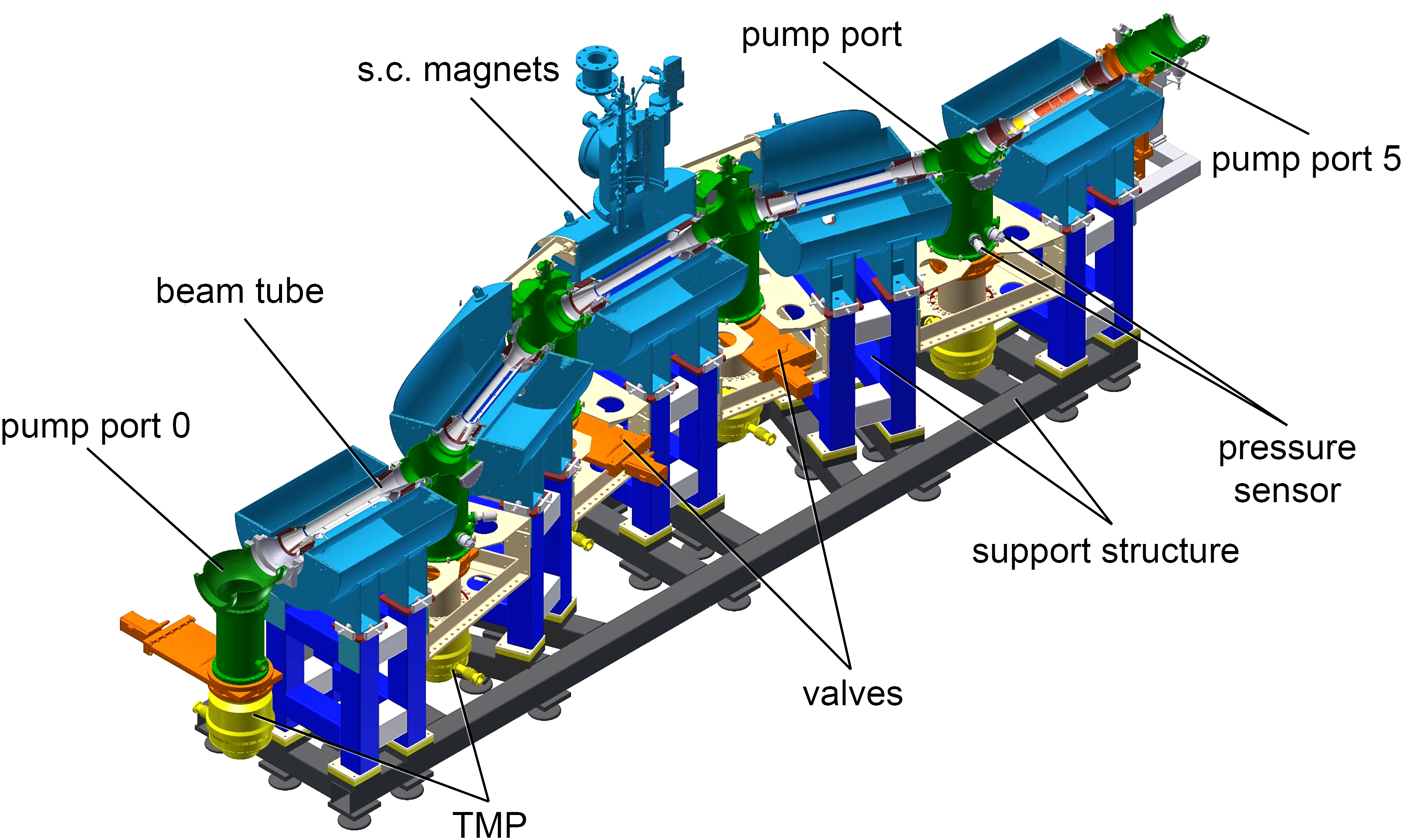}
	\end{center}
	\caption{\label{fig:dps_cad}Shown is the modular setup of the DPS2-F as a CAD drawing with a half-cut along the horizontal plane. Each beam tube (silver) is located in a warm bore of a superconducting solenoid (blue). The beam tubes are connected via the pump ports (green) to one beamline. The valves (orange) are located between the pump ports and the entrance of the TMPs (yellow). }
\end{figure}

The DPS2-F has five pump ports (PP0, PP1, ..., PP4), which are aligned perpendicular to the interconnecting 
beamlines (BT1-5). Pump port 0 was added at a late stage during the design of the pumping section, 
which is why in previous publications the DPS2-F is mostly described with only four pump ports~\cite{Luo2006, Malyshev2007}. 
Incoming and outgoing beamlines of PP1-4 change direction at the pump ports by 20$^\circ$ 
(see Fig.~\ref{fig:dps_cad}). This geometry prevents a direct line of sight and increases the 
number of collisions with the walls, and thus, the pumping probability for the neutral molecules. 
Electrostatic dipoles (half-pipe-shaped stainless-steel electrodes) and ring-shaped electrodes 
remove tritium ions by either drifting them towards the walls where they are neutralized, or 
reflecting them back towards the source. 

The beamline of the DPS-2F, between the WGTS and the CPS cryostat, is 7.3-m long. Each beam tube consists of a central tube, connected via bellows to a flange on each side. The central tubes have an inner diameter of \SI{100}{mm}. The smallest diameter of the beamline is $\mathrm{85\,mm}$, defined by the beamline instrumentation. The entire DPS2-F has a weight of about $\mathrm{10^{4}\,kg}$ and is fixed to the floor with earthquake protections. A computer-aided design (CAD) drawing of the DPS2-F is shown in Fig.~\ref{fig:dps_cad}.

\subsubsection{Vacuum system}
The required gas flow reduction factor is achieved by six TMPs (Leybold MAG-W 2800). Two TMPs are 
connected to PP0. The other four TMPs are located at the lower ends of PP1-4. Each TMP can be  
separated from the beamline by a DN 250\,mm gate valve (VAT series 10). UHV vacuum gauges 
(MKS 421 inverted magnetron) are mounted on each pump port. The pressure along the DPS2-F beamline 
from PP0 to PP4 drops from approximately $10^{-6}$\,mbar to $10^{-8}$\,mbar. This absolute 
pressure reduction is not to be confused with the reduction of the tritium flow and the associated 
reduction of the partial pressure simulated here. 

The superconducting magnets surrounding the beamline provide a guiding field of up to 5.5\,T~\cite{arXivArenz2018e}. 
A passive magnetic shield encases each TMP to prevent eddy currents from heating up, and possible crashing, of the fast-moving rotors~\cite{Wolf2011}.

\subsection{The Cryogenic Pumping Section}

The last part of the transport and pumping section is the Cryogenic Pumping Section (CPS), which has to reduce the residual gas flow by more than seven orders of magnitude. 
For this purpose, a cold argon frost layer prepared on the inner beamline surface, with an area of about \SI{2}{\meter\squared} that is maintained at 3 K, adsorbs the incoming tritium molecules.

\subsubsection{Geometry}

\begin{figure}
	\begin{center}
		\includegraphics[width=\linewidth]{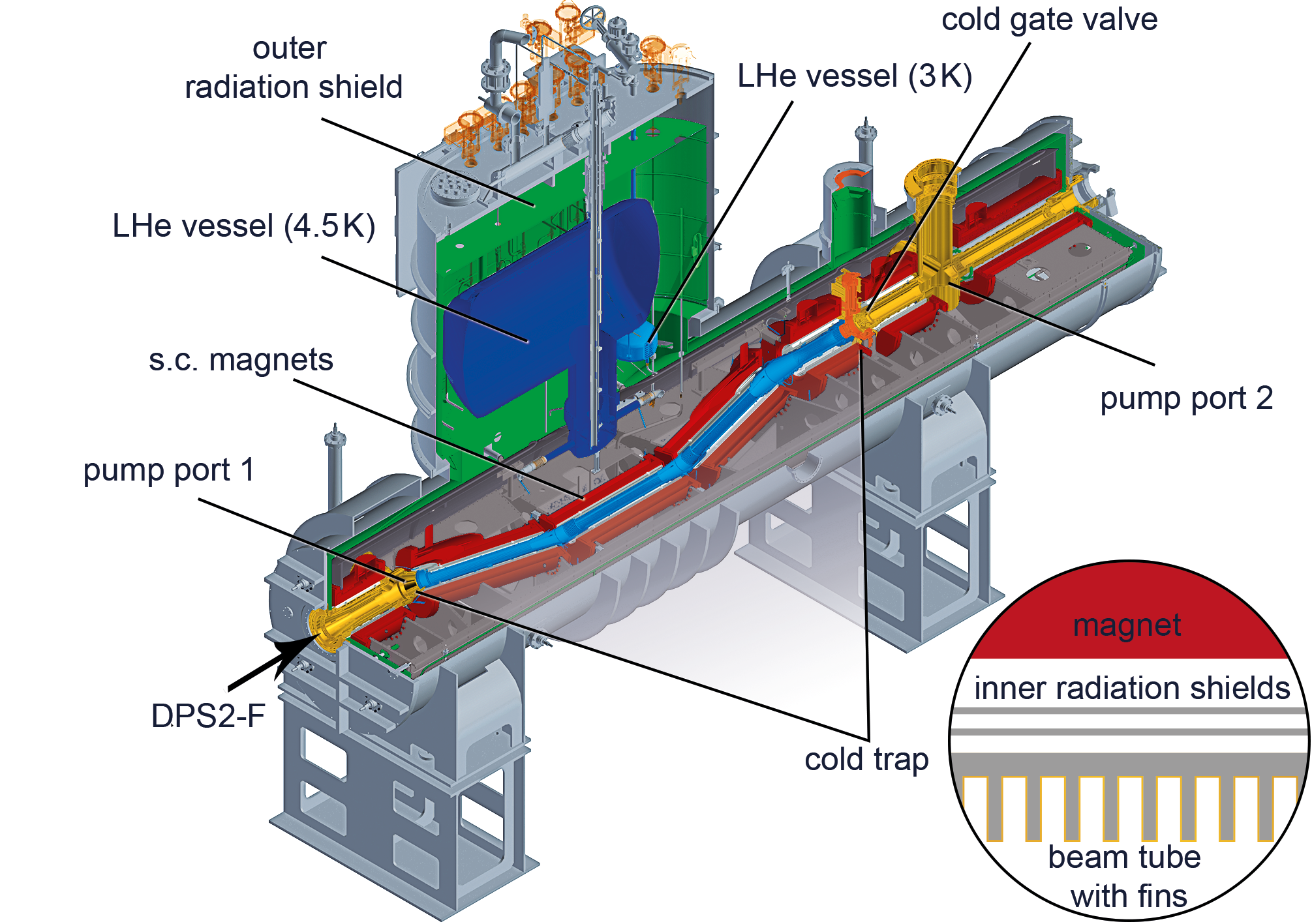}
	\end{center}
	\caption{\label{fig:cps_cad}CAD drawing of the CPS cryostat in 3/4 section. The gold-plated beam tube is	surrounded by seven superconducting magnets (in red). The LHe vessel (\SI{4.5}{\kelvin}) provides a reservoir of \SI{4.5}{\kelvin} cold helium, which is used for the cooling of the magnets and beam tube. The cold trap can be seen in blue between pump port 1 and the cold gate valve (CGV). The CGV is a safety valve operated inside the CPS at a temperature of \SI{4.5}{\kelvin}. The differential pumping section (DPS2-F) is connected on the left side, the Pre-Spectrometer on the right side.}
\end{figure}

In Fig.~\ref{fig:cps_cad} a CAD drawing of the CPS is shown.
The 12-ton CPS cryostat built by ASG Superconductors S.p.A. is about \SI{6.5}{\meter} long and \SI{4}{\meter} high.
The beam tube elements of the CPS are subdivided into seven sections with a total length of roughly \SI{7}{m} (inner diameter: $\ge$\SI{75}{mm}) and two pump ports. 
Each section is surrounded by a superconducting solenoid that produces the \SI{5.6}{\tesla} magnetic field~\cite{arXivArenz2018e} to guide $\upbeta$-electrons adiabatically through the CPS.
The second and fourth beam tube are rotated by 15$^\circ$ from the longitudinal spectrometer axis, so that neutral tritium molecules would hit the beam tube wall, where they are adsorbed.
Each beam tube is a stainless steel tube with gold plated on the inner surface.
Additionally, there are 90 circular fins (length of \SI{2}{mm}) inside each of the beam tube sections 2-4 enlarging the inner surface~\cite{Gil2010}.

\subsubsection{Vacuum system}

The main part of the vacuum system of the CPS is the 3-meter long cold trap (sections 2-5), covered by an argon frost layer at \SI{3}{\kelvin}.
The gold plated inner surface provides a clean surface for argon frost crystallization, as well as reducing the diffusion of hydrogen isotopes into the stainless steel.
In order to reach \SI{3}{\kelvin} in the LHe vessel (\SI{3}{\kelvin}) the \SI{4.5}{\kelvin} helium is pumped down to a pressure of \SI{0.16}{\bar} and circulated through the cooling loop of the cold trap.
For safety reasons the argon layer will be regenerated after 60 days of measurement time, which corresponds to an accumulated tritium activity below \SI{1}{Ci} stored on the cold trap.

The other beam tube elements are cooled with liquid nitrogen and therefore operated at about \SI{77}{\kelvin}.
At PP1 and PP2 there are cold cathodes (MKS 421 inverted magnetron) to monitor the pressure in front of and behind the cold trap.
TMPs are installed to both pump ports, but are turned off during standard KATRIN operation.

\subsubsection{Cold trap temperature profile}
\label{sub:beamline_temp}

The tritium molecules adsorbed on the frost layer can redesorb either via beta decay or by thermal desorption.
Therefore the temperature of the argon frost layer is an important parameter for the pumping efficiency of the CPS.
The correlation of the mean sojourn time $\tau_{\mathrm{des}}$ and the temperature $T$ is~\cite{lit:jousten}
\begin{equation}
	\label{eq:tau_des}
	\tau_{\mathrm{des}}=\tau_0\cdot\exp(\frac{E_{\mathrm{des}}}{RT}) \ ,
\end{equation}
where $\tau_0$ is the adsorbed particle's period of oscillation perpendicular to the surface, $E_{\mathrm{des}}$ is the desorption energy for one mole of adsorbed gas and ${R=\SI{8.314}{\joule\mole\per\kelvin}}$ the molar gas constant.

For monitoring the temperature three rhodium-iron sensors with \SI{50}{\milli\kelvin} accuracy are attached to each beam tube section of the cold trap.
During the first activation of the \SI{3}{\kelvin}-cooling system, the measured temperatures on the beam tube did not reach the expected \SI{3}{\kelvin}~\cite{Roettele2017}, but varied between \SI{3.4}{\kelvin} and \SI{6.2}{\kelvin}.

In order to investigate the origin of the temperature discrepancy, the heat transfer module of the commercial simulation program COMSOL Multiphysics\textsuperscript{\textregistered} was used with a finite-element-method simulation.
A CAD model of the cold trap geometry is imported; the model includes the magnetic coils 2-5, the inner radiation shields, part of the \SI{3}{\kelvin}-cooling loop connected to the beam tube, and the beam tube.
The simulation is initialized with a fixed temperature of \SI{4.5}{\kelvin} for the magnets and \SI{3}{\kelvin} for the pipes of the cooling loops around the beamline.
To reduce the calculation time, radiation is only allowed between opposite surfaces, e.g. between magnetic coils and inner radiation shield, while the radiation inside the beam tube is neglected.
The heat load from elements, which are not explicitly simulated in the geometry model, is taken into account by assuming a uniform thermal black body radiation with a specific temperature.
In COMSOL Multiphysics\textsuperscript{\textregistered} this parameter is called ambient temperature.
In order to minimize the differences between the simulated and the measured temperatures, the ambient temperatures for the different beam tube sections vary between 70 and \SI{90}{\kelvin}.
In this way the non-negligible radiation of the pump ports is included.

\begin{figure}
	\begin{center}
		\includegraphics[width=\linewidth]{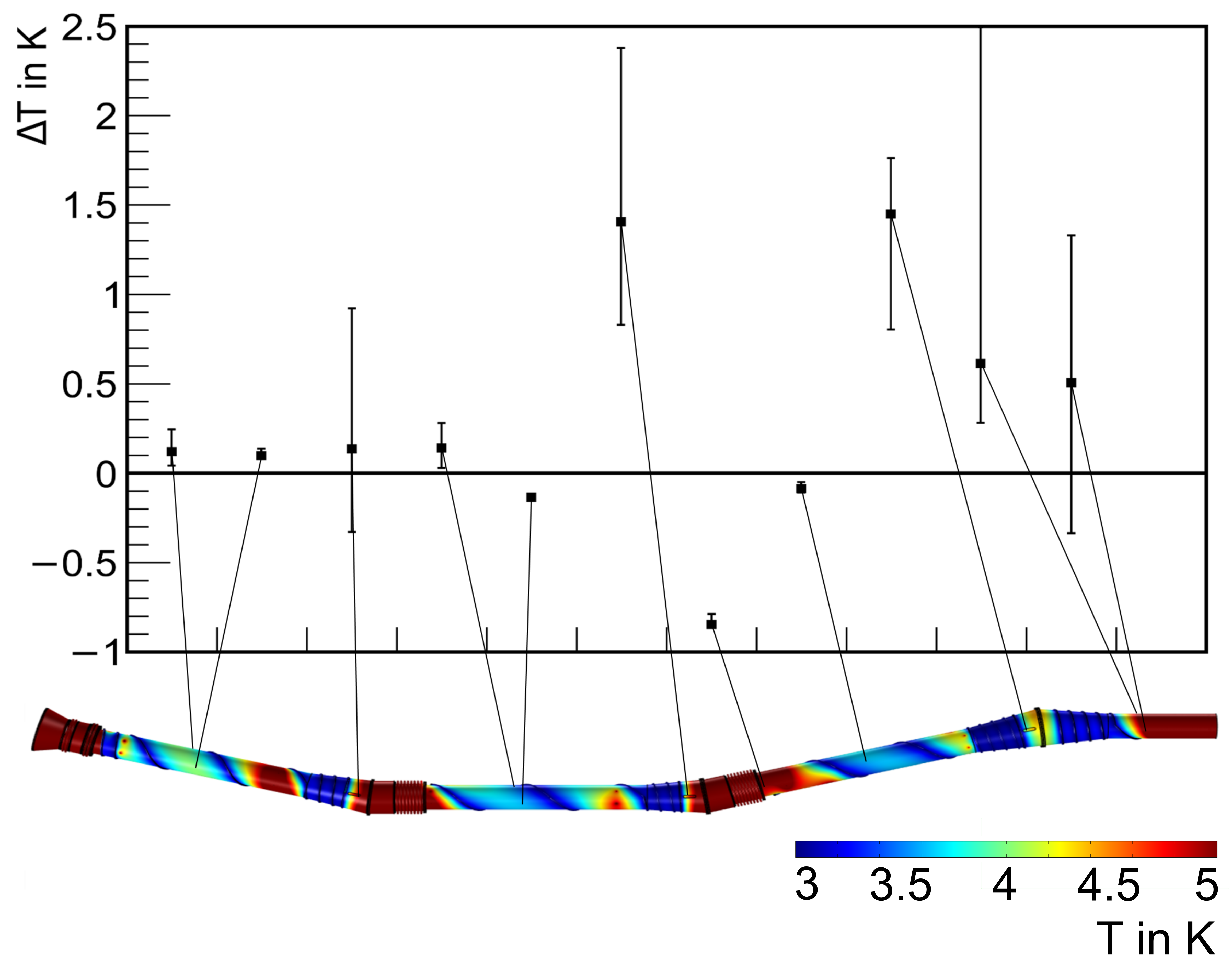}
	\end{center}
	\caption{\label{fig:temp_dev}Temperature deviation in Kelvin between measurement ($T_{\mathrm{meas}}$) and simulation ($T_{\mathrm{sim}}$). The error bars correspond to the simulated temperature variation across the connecting area of the sensor housing on the beam tube. At the bottom the simulated COMSOL Multiphysics\textsuperscript{\textregistered} temperature profile is shown. The lines connect the $\Delta T=T_{\mathrm{meas}}-T_{\mathrm{sim}}$ points to the positions of the sensors.}

\end{figure}

In Fig.~\ref{fig:temp_dev}, the deviations to the measurement results are shown.
The errors correspond to the temperature gradient along the $\SI{30.0}{\milli\meter}\times\SI{14.8}{\milli\meter}\times\SI{19.5}{\milli\meter}$ copper sensor housing connected to the beamline.
The largest discrepancies are located in the regions near to the beam tube bends and in beam tube section 5.
The first one can be explained by the higher radiative heat loads on these regions due to the gaps between the inner radiation shields and the magnets.
In beam tube section 5, a \SI{180}{mm} stretch of the cooling loop is not brazed to the beam tube due to a repair in this area, which leads to a large area on the right side that has a temperature above \SI{5}{\kelvin}.
Regions near the cooling loop reach the expected \SI{3}{\kelvin} while the temperature increases further away.
Due to the narrower windings of the cooling loop towards the end of the beam tube sections, there are areas which are homogeneous at \SI{3}{\kelvin}.
Hot spots arise as a result of bolts connecting the warmer inner radiation shield to the beam tube sections 2-4.
Except for these hot spots, most of the beamline areas, in particular those with the fins, are in a temperature range between 3 and \SI{4}{\kelvin}.
This simulated temperature profile is used in the next sections to calculate the reduction factor of the cold trap.

	\section{The TPMC Models \label{sec:TPMC_models}}
\label{sec:tpmcmodel}

The TPMC models have been simulated with the software package MolFlow+ (version 2.5.6)~\cite{Kersevan2009}. The software is designed for particle tracking in the molecular flow regime. Its basic concepts are described in the following section.
 
\subsection{Simulation with MolFlow+}

MolFlow+ tracks test particles through the model of a vacuum chamber build up by a polygon mesh, of so-called facets. The particles only interact with the walls. When they hit a facet, they can either be adsorbed, reflected or transmitted. The properties of a facet are defined by various parameters, such as the temperature, the sticking factor defining the adsorption probability, the type of reflection, and the opacity. A facet can also be defined as a desorbing source of gas, where new particle tracks originate. Each track is simulated through a series of diffuse reflections and transmissions at facets until it is finally adsorbed. Each facet has three counters that are assigned for desorptions, hits and adsorptions; the counters are incremented accordingly when hit by a particle. A pump, such as a TMP, is represented by a circular facet the size of its entrance flange and a sticking factor $\alpha \in [0,1]$ equaling its pumping probability for the simulated gas type. 
If a particle is reflected, we assume diffuse reflection, following the cosine law. Apart from the opaque facets representing the walls and internal structures of the vacuum chamber, the user can also define virtual facets, which are transparent. They do not affect the path of a particle, but count its transmission as a hit. A one-sided facet counts only particles impinging from one direction, while a two-sided one counts all particles. Virtual facets can be placed anywhere in the chamber, allowing us to monitor the pressure in the volume or to determine the transmission of particles from one part of the model to another.

After the simulation of an appropriate number of particle tracks, the results, represented by the three counters of each facet, are in general used to determine conductances, effective pumping speeds, and partial pressures.  Details of the analysis methods for the DPS2\nobreakdash-F and the CPS are described in the following sections.

\subsection{The DPS2-F Model}
\label{sec:dpsmodel}
\begin{figure}[]
	\centering
	\includegraphics[width=\linewidth]{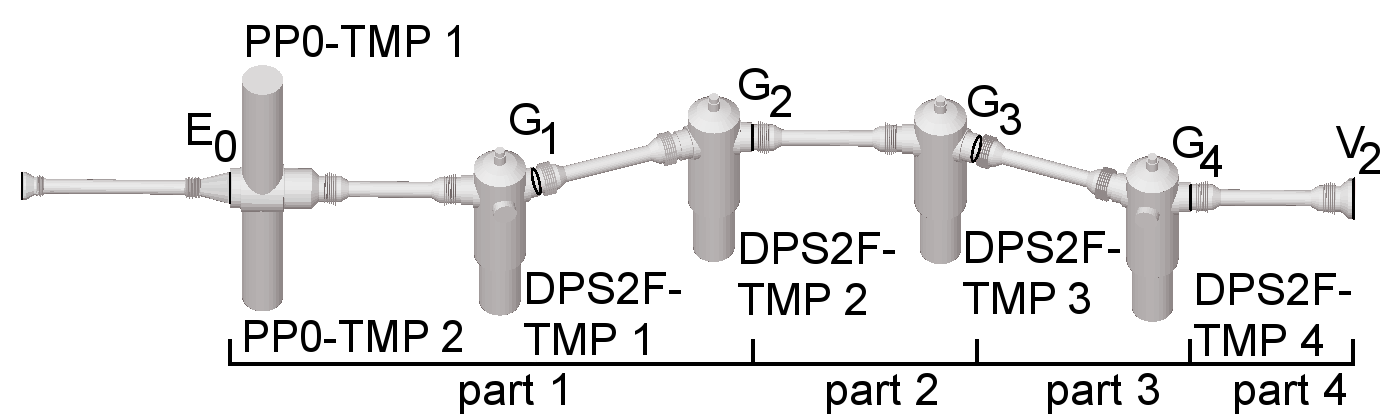}
	\caption{The MolFlow+ model of the DPS2\nobreakdash-F vacuum system. Some virtual facets, valves and the six active turbomolecular pumps are denoted. The four concatenated parts are indicated. }
	\label{fig:dps-model}
\end{figure}

For DPS2\nobreakdash-F particle tracking simulations, the MolFlow+ model shown in Fig.~\ref{fig:dps-model} is used~\cite{PhDJansen2015}. It comprises about 19000 facets. The models of the WGTS~\cite{arXivKuckert2018} and the DPS2\nobreakdash-F join at the entrance of PP0. 
The TMP sticking factors $\alpha=0.252$ correspond to the pumping speed of TURBOVAC MAG W 2800 pumps for a gas mass of $m=\SI{6}{\gram\per\mole}$, which has been chosen as the particle mass for the simulation of tritium molecules. It should be mentioned that the main systematic uncertainty of the simulations is the pumping probability $\alpha$; the uncertainty is estimated using an empirical model for the pumping probability by Malyshev~\cite{Malyshev2007} to be 20\%. 
The geometry of the electrodes of the ion detection and removal system inside the beam tube is included.

The particles desorb from the entrance facet of PP0 and are tracked through the complete geometry until they get pumped by one of the six TMPs or reach the entrance valve V2 towards the CPS. A sticking factor of $\alpha=1.0$ has been assigned to the V2 facet. The entrance facet of PP0 is transparent.
On the upstream side the geometry is terminated by the last beam tube element of DPS1-F, which ends in a facet with sticking factor $\alpha=1.0$, representing the TMPs of the second stage of the DPS1-F.

\subsubsection{Concatenation of subsequent sections}
\label{ch:concatenation}

The simulation of a flow reduction factor ${R = \Phi_{\mathrm{in}}/\Phi_{\mathrm{out}}}$ of 5 orders of magnitude and more is very time consuming. Therefore, the geometry of the DPS2\nobreakdash-F was subdivided into four different parts (see Fig.~\ref{fig:dps-model}). Each part was simulated independently, and the individual results were concatenated subsequently to derive the transmission probability, hits (for pressure) and adsorptions~\cite{PhDJansen2015}. The reduction factor can be calculated by dividing the number of started particle tracks $N_{\mathrm{des,E}_0}$ at facet $\mathrm{E}_0$ of PP0 by the number of particles $N_{\mathrm{ads,V}_2}$ adsorbed (leaving DPS2\nobreakdash-F) at the gate valve $\mathrm{V}_2$ between the DPS2\nobreakdash-F and the CPS:
\begin{equation}
\begin{split}
R_{\rm tot} &= R_{1} \cdot R_{2} \cdot R_{3} \cdot R_{4} \\
&= \left(\frac{N_{\mathrm{des,E}_0}}{N_{\mathrm{hit,G}_2}} \right)_{\mathrm{part \ 1}} \cdot \left(\frac{N_{\mathrm{hit,G}_2}}{N_{\mathrm{hit,G}_3}} \right)_{\mathrm{part \ 2}} \cdot \left(\frac{N_{\mathrm{hit,G}_3}}{N_{\mathrm{hit,G}_4}} \right)_{\mathrm{part \ 3}}\\
& \ \ \ \cdot \left(\frac{N_{\mathrm{hit,G}_4}}{N_{\mathrm{ads,V}_2}} \right)_{\mathrm{part \ 4}}.
\end{split}
\end{equation}  
Here $N_{\mathrm{hit,G}_{2-4}}$ are the number of particles passing the virtual facets in the downstream direction between the parts indicated in Fig.~\ref{fig:dps-model}.
Apart from the first part, the particle tracks were started at the entrances G$_{1-3}$ of the preceding beam tube section. Thus, the resulting solid-angle distribution of the velocities of the incoming tracks at facets G$_{2-4}$ was closer to the one expected for a single-pass simulation of the entire geometry. All simulations were done with the full model, changing only the desorbing facet where the tracks started. The simulations ran until the statistical uncertainty of the hits in the respective concatenating facets was better than 1\%.

\subsection{The CPS Model}
\label{ch:CPS-TPMC-model}
\begin{figure}[]
	\centering
	\includegraphics[width=\linewidth]{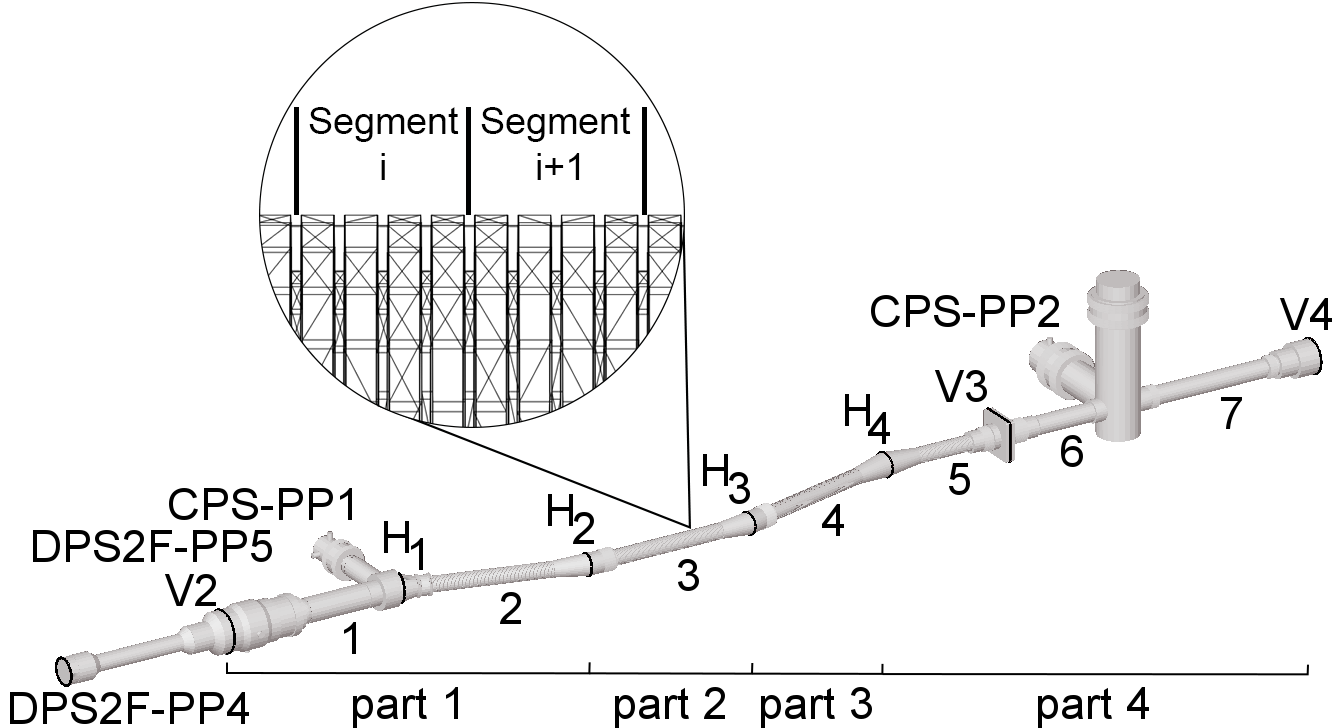}
	\caption{The MolFlow+ model of the CPS vacuum system. Some virtual facets and real valves are marked by black lines. The four parts which are concatenated are marked. }
	\label{fig:cps-model}
\end{figure}

The MolFlow+ model of the CPS is shown in Fig.~\ref{fig:cps-model}. It consists of about 58000 facets. The geometry of the CPS starts at valve V2 and ends at valve V4. Seven elements build up the complete beam tube. The TMPs at the two pump ports are turned off during standard operation, so the only pumping mechanism is cryosorption. For particles reflected back into the DPS2\nobreakdash-F the last beam tube element is included up to the last DPS2\nobreakdash-F pump port (PP4). Particles reaching this pump port are considered to be pumped out with high probability ($\alpha = 1$). This scheme ensures that the concatenation of the DPS2\nobreakdash-F and CPS models is simulated with matching boundary conditions. A particle leaving the DPS2\nobreakdash-F in the previous simulation through valve V2 is adsorbed ($\alpha = 1$), not taking into account that a fraction of the particles is actually reflected back into the DPS2\nobreakdash-F. This is done in the subsequent CPS simulation, making sure that the back-reflection is not taken into account twice. 

In order to reduce the computing time, the CPS model is subdivided into four parts, which are simulated separately and concatenated afterwards, similar to the DPS2\nobreakdash-F (see Sec.~\ref{ch:concatenation}). Since the sticking factor of the argon frost layer is not precisely known and also depends on the initial coverage, simulations were performed for $\alpha = 0.0$ to 0.7 in steps of 0.1. The upper value of 0.7 is the expected sticking factor for a well prepared argon layer at 3\,K~\cite{Yuferov1970}. 
For facets not belonging to the cold trap, the sticking factor is set to $\alpha=0.0$. At both ends of the model (DPS2F\nobreakdash-PP4 and V4) it is set to $\alpha = 1$. In this case, the particles are either pumped out at DPS2F\nobreakdash-PP4 or enter the Pre-Spectrometer volume, which has a diameter approximately twenty times larger than the CPS beam tube and a more than two orders of magnitude higher effective  pumping speed for tritium than the conductance between the Pre-Spectrometer and the cold section of the CPS. 

The tracking of gas particles starts at valve V2. Before a particle reaches the cold trap it is 
only diffusely reflected when hitting a wall. Once a particle is adsorbed on a facet of the cold 
trap it sticks on it forever and the tracking path in MolFlow+ ends. 
According to Eq.~\ref{eq:tau_des} in 
Sec.~\ref{sub:beamline_temp} this model is not appropriate since particles can leave the facet by 
thermal desorption after some characteristic sojourn time $\tau_{\rm des}$. The redesorbed gas 
slowly migrates towards the end of the CPS, adding to the gas leaving through valve V4 towards 
the Pre-Spectrometer. Since the incoming gas flow from the DPS2\nobreakdash-F is assumed to be constant, the 
reduction factor $R_{\mathrm{CPS}} = \Phi_{\mathrm{in}}/\Phi_{\mathrm{out}}$ of the CPS decreases 
over time.

Therefore a model has been developed, which combines the results from a multitude of MolFlow+ 
simulations and, in a second step, adds the effect of a 
finite sojourn time $\tau_{\rm des}$ on the adsorbed tritium 
molecules. The basic idea is to subdivide the cold trap of the CPS beam tube into $n = 102$ 
smaller segments, and consider each segment as an individual cryo pump, where particles can 
adsorb and redesorb.

\subsubsection{Segmentation of the cold trap}

The number of adsorbed particles $A_i(t)$ sitting on the surface of segment $i$ can either 
originate from the incoming gas flow $\Phi_\mathrm{in}$ through the DPS-2F or from the 
desorption off 
other segments $j$. The change in the number of particles adsorbed on segment $i$ can be 
described by a system of coupled differential equations

\begin{equation}
\frac{\mathrm{d}A_i(t)}{\mathrm{d}t} = \Phi_{\mathrm{in}} \cdot U_{\rm{des,V2}}^{\mathrm{ads,}i} 
+ \sum_{j=1}^{n} \left( \frac{A_j(t)}{\tau_{\mathrm{des}}}\cdot 
U_{\mathrm{des,}j}^{\mathrm{ads,}i} \right)-  \frac{A_i(t)}{\tau_{\rm{des}}} \ . \label{eqn:DGL}
\end{equation}
The first term describes the adsorption rate on segment $i$ from $\Phi_\mathrm{in}$. The 
parameter $U_{\rm{des,V2}}^{\mathrm{ads,}i}$ is the adsorption probability on segment $i$ for 
particles entering the CPS through valve V2. The second term is the sum of adsorptions on 
segment $i$ of particles desorbing from all segments $j$ of the cold trap. The desorption rate 
$A_j(t)/\tau_{\mathrm{des}}$ is proportional to the number of adsorbed particles $A_j(t)$ 
and the adsorption probability $U_{\mathrm{des,}j}^{\mathrm{ads,}i}$. The third 
term subtracts the rate of desorbing particles from segment $i$.  
All adsorption probabilities can be combined in a matrix $U$ determined by MolFlow+ simulations. 

The segments in the cylindrical part of the beam tubes were chosen to be \SI{3.2}{\centi\meter} long, which corresponds to the distance between four consecutive fins (see Fig.~\ref{fig:cps-model}). The cones at the ends of each beam tube element were simulated as longer segments. In addition, four simulations were necessary for simulating the gas inlet at valve V2, and the concatenations of the four parts, similar to the DPS2\nobreakdash-F, for calculating the direct transmission through the CPS from V2 to V4. For the direct gas flow through V2, the results from these concatenating simulations were also used to calculate the adsorption probabilities $U_{\rm{des,V2}}^{\mathrm{ads,}i}$ for segments beyond H2 (see Fig.~\ref{fig:cps-model}). For each segment $i$ the counters of the corresponding facets were added up to the number of desorptions $N_{\mathrm{des,}i}$, adsorptions $N_{\mathrm{ads,}i}$ and hits $N_{\mathrm{hit,}i}$. With these numbers the probability matrices for desorptions ($U$) and pressure (hit matrix $V$) can be calculated:

\begin{itemize}
\item \textbf{$U_{\mathrm{des,}j}^{\mathrm{ads,}i}=\frac{N_{\mathrm{ads,}i}}{N_{\mathrm{des,}j}}$} is the probability that a particle desorbing from segment $j$ is adsorbed on segment $i$. 
\item \textbf{$V_{\mathrm{des,}j}^{\mathrm{hit,}i}=\frac{N_{\mathrm{hit,}i}}{N_{\mathrm{des,}j}}$} is the probability that a particle desorbing from segment $j$ hits segment $i$. 
\end{itemize}
It is $j \in [0,102]$ and $i \in [1,103]$ for $U_{\mathrm{des,}j}^{\mathrm{ads,}i}$ where $j=0$ represents the inlet valve V2, $i,j \in [1,102]$ represent the 102 segments, and $i=103$ the exit valve V4. For $V_{\mathrm{des,}j}^{\mathrm{hit,}i}$ it is $j \in [0,102]$ and $i \in [1,105]$. The indices $i=104$ and $i=105$ represent the facets where the pressure gauges are located in CPS-PP1 and CPS-PP2, respectively.

In order to attain better statistics in the region, where most of the gas is adsorbed, the concatenations of the CPS geometry were also applied for calculating the hit $V_{\mathrm{ads,}i}^{\mathrm{des,}j}$ and adsorption  $U_{\mathrm{ads,}i}^{\mathrm{des,}j}$ probabilities for desorptions from the segments of the first cold trap section ($j \in [1,27]$):

\begin{enumerate}[i)]
\item If the segments $i$ and $j$ lie in the same or in the neighboring beam tube section it is: 
\\$U_{\text{des,}j}^{\text{ads,}i}=\frac{N_{\text{ads,}i}}{N_{\text{des,}j}}$ and $V_{\text{des,}j}^{\text{hit,}i}=\frac{N_{\text{hit,}i}}{N_{\text{des,}j}}$.
\item If the segments $i$ and $j$ are separated by exactly one beam tube section with the ending facet $H_3$ it is: 
\\$U_{\text{des,}j}^{\text{ads,}i}=\frac{N_{\text{hit,H}_3}}{N_{\text{des,}j}} \cdot\left( \frac{N_{\text{ads,}i}}{N_{\text{hit,H}_3}} \right)_{\mathrm{part \ 3}}$ and \\ $V_{\text{des,}j}^{\text{hit,}i}=\frac{N_{\text{hit,H}_3}}{N_{\text{des,}j}} \cdot \left( \frac{N_{\text{hit,}i}}{N_{\text{hit,H}_3}} \right)_{\mathrm{part \ 3}}$.
\item If the segments $i$ and $j$ are separated by exactly two beam tube sections with the ending facets $H_3$ and $H_4$ it is: 
\\$U_{\text{des,}j}^{\text{ads,}i}=\frac{N_{\text{hit,H}_3}}{N_{\text{des,}j}} \cdot\left( \frac{N_{\text{hit,H}_4}}{N_{\text{hit,H}_3}} \right)_{\mathrm{part \ 3}}  \cdot\left( \frac{N_{\text{ads,}i}}{N_{\text{hit,H}_4}} \right)_{\mathrm{part \ 4}}$ and\\ $V_{\text{des,}j}^{\text{hit,}i}=\frac{N_{\text{hit,H}_3}}{N_{\text{des,}j}}\cdot\left( \frac{N_{\text{hit,H}_4}}{N_{\text{hit,H}_3}} \right)_{\mathrm{part \ 3}} \cdot\left( \frac{N_{\text{hit,}i}}{N_{\text{hit,H}_4}} \right)_{\mathrm{part \ 4}}$.
\end{enumerate} 
This case-by-case analysis was constructed in such a way that the solid angle under which a particle enters the next part of the CPS is comparable to a single-pass simulation. 
For elements $j \in [28,102]$, the concatenation has not been applied since the first test simulations indicated a coverage reduced by several orders of magnitude compared to the first CPS simulation part.  

\subsubsection{Time-dependent gas flow}

In order to describe time-dependent processes the coupled differential equations~\ref{eqn:DGL} of the amounts of adsorbed gas on each segment are numerically integrated over time with discrete steps of $\Delta t$. The gas inlet into the CPS starts at $t_0=0 \,$s. The inlet rate $\Phi_{\mathrm{in}}$ from the DPS2\nobreakdash-F stays constant over the whole time. 
In the simulation, this gas inlet is represented by the desorption of particles from facet V2. 
The time difference between $t_0$ and any other time $t_n$ is subdivided into $n$ intervals with length $\Delta t$. At the time $t_0=0 \,$s, it is assumed that there are no particles in the system at all ($A_i(0) = 0$), which defines the boundary conditions for the numerical integration. After the first iteration at $t_1 = \Delta t$ the amount of gas adsorbed on a specific segment $i$, is described by 

\begin{equation}
A_i(t_1) =\Phi_{\mathrm{in}} \cdot \Delta t \cdot U_{\rm{des,V2}}^{\mathrm{ads,}i}.
\end{equation} 
For any time $t_n=n \cdot \Delta t>t_1$ the number of adsorbed particles on segment $i$ is defined as

\begin{equation}
\begin{split}
A_i(t_n) &= A_i(t_{n-1}) + \Phi_{\mathrm{in}} \cdot \Delta t \cdot U_{\rm{des,V2}}^{\mathrm{ads,}i} \\
&+ \sum_{j=1}^{102} \left(A_j(t_{n-1}) \frac{\Delta t}{\tau_{\rm{des}}} U_{\mathrm{des,}j}^{\mathrm{ads,}i} \right)- A_i(t_{n-1}) \frac{\Delta t}{\tau_{\rm{des}}}
\end{split}
\end{equation}
for $i \in [1,102]$.

The first term represents the number of adsorbed particles at the time $t_{n-1}$. The second term takes the gas inlet between $t_{n-1}$ and $t_{n}$ into account. The factor $A_j(t_{n-1}) \Delta t/\tau_{\rm{des}}$ in the third term equals the amount of desorbed gas from segment $j$. By multiplying this with $U_{\mathrm{des,}j}^{\mathrm{ads,}i}$, one gets the number of particles desorbed from segment $j$ and adsorbed on segment $i$ during the time interval $\Delta t$. For the gas $\Phi_{\mathrm{out}}$ leaving the CPS towards the spectrometer section through valve V4 ($i=103$), we neglect any gas coming back to the CPS due to the large pumping speed of the Pre-Spectrometer. With 

\begin{equation}
\begin{split}
\Phi_{\mathrm{out}} &= \frac{A_{103}(t_n)-A_{103}(t_{n-1})}{\Delta t} \\
 &= \Phi_{\mathrm{in}} \cdot U_{\rm{des,V2}}^{\rm{ads,103}} + \sum_{j=1}^{102} \left( \frac{A_j(t_{n-1})}{\tau_{\rm{des}}} U_{\mathrm{des,}j}^{\rm{ads,103}} \right) \ ,
\end{split}
\end{equation}
we can describe the time-dependent flow reduction as

\begin{equation}
R(t_n)=\frac{\Phi_{\mathrm{in}}}{\Phi_{\mathrm{out}}}=\frac{\Phi_{\mathrm{in}}}{\Phi_{\mathrm{in}} \cdot U_{\rm{des,V2}}^{\rm{ads,103}}+\sum_{j=1}^{102} \left(\frac{A_j(t_{n-1})}{\tau_{\rm{des}}} U_{\mathrm{des,}j}^{\rm{ads,103}} \right)} \ .
\end{equation}
The pressure at segment $i$ in the TPMC simulation can be determined from the number of hits $N_{\mathrm{hit}, i}$, the area of the segment $F_i$, the mean thermal velocity $\bar{c}$, the number of desorbed particles $N_{\mathrm{des}, j}$ from segment $j$, and the actual gas flow (or outgassing rate) into the chamber $Q_j(t)$~\cite{Kersevan2014}:

\begin{equation}
p_{ij}(t)=\frac{4}{\bar{c}} \frac{Q_j(t)}{F_i} \frac{N_{\mathrm{hit,}i}}{N_{\mathrm{des,}j}}=\frac{4}{\bar{c}} \frac{Q_j(t)}{F_i} \cdot V_{\mathrm{des,}j}^{\mathrm{hit,}i}.
\end{equation}
Multiplying the particle flow $\Phi_{\mathrm{in}}$ with the Boltzmann constant $k_{\rm B}$ and the temperature $T$, the pressure after the first iteration at $t_1 = \Delta t$ is

\begin{equation}
p_i(t_1) =\frac{4}{\bar{c}} \frac{\Phi_{\mathrm{in}} \cdot k_BT}{F_i} \cdot V_{\rm{des,V2}}^{\mathrm{hit,}i} \ ,
\end{equation}
and for $t_n=n \cdot \Delta t$:

\begin{equation}
\begin{split}
p_i(t_n) &=p_i(t_1)+ \sum_{j=1}^{102} p_{ij} \\
&= p_i(t_1) + \sum_{j=1}^{102} \left(\frac{4}{\bar{c}} \frac{A_j(t_{n-1}) \cdot k_B T}{F_i \cdot \tau_{\rm{des}}} V_{\mathrm{des,}j}^{\mathrm{hit,}i}\right) \\
&=\frac{4 \cdot k_BT}{\bar{c} \cdot F_i} \left(\Phi_{\mathrm{in}} \cdot V_{\rm{des,V2}}^{\mathrm{hit,}i}+\sum_{j=1}^{102} \left( \frac{A_j(t_{n-1})}{\tau_{\rm{des}}} V_{\mathrm{des,}j}^{\mathrm{hit,}i}\right)\right).
\end{split}
\end{equation}
The pressures at CPS-PP1 and CPS-PP2 are particularly relevant since these are measurable quantities. From simulations, the pressure ratio is

\begin{equation}
\frac{p_{\text{PP1}}(t_n)}{p_{\text{PP2}}(t_n)}
=\frac{\frac{\Phi_{\mathrm{in}}}{F_{\text{PP1}}} \cdot V_{\text{des,V2}}^{\text{hit,PP1}} + \sum_{j=1}^{102} \frac{A_j(t_{n-1})}{\tau_{\text{des}}} \frac{1}{F_{\text{PP1}}} V_{\text{des,}j}^{\text{hit,PP1}}}{\frac{\Phi_{\mathrm{in}}}{F_{\text{PP2}}} \cdot V_{\text{des,V2}}^{\text{hit,PP2}} + \sum_{j=1}^{102} \frac{A_j(t_{n-1})}{\tau_{\text{des}}} \frac{1}{F_{\text{PP2}}} V_{\text{des,}j}^{\text{hit,PP2}}} .
\end{equation}
This ratio is correlated with the flow reduction:

\begin{equation}
\label{eq:reduction-pressure}
R(t_n)=k(t_n) \cdot \frac{p_{\text{PP1}}(t_n)}{p_{\text{PP2}}(t_n)}. 
\end{equation}
By simulating both the flow reduction and the pressure ratio, the ad hoc factor $k(t_n)$ can thus be determined. This result is essential for interpreting measured pressure values at the pump ports. 

\subsubsection{Including the beamline temperature profile}
\label{sub:includetempprofile}

The COMSOL Multiphysics\textsuperscript{\textregistered} simulation of the CPS cold trap temperature profile in Sec.~\ref{sub:beamline_temp} revealed inhomogeneities of several Kelvin. These inhomogeneities have a non-negligible influence on the pumping efficiency of the cold trap; in particular, the mean sojourn time $\tau_{\rm des}$ is affected. This can be included in the analysis of the MolFlow+ simulations by calculating the weighted mean of all $\bar{\tau}_{\rm des}$ in the system:
\begin{equation}
\begin{split}
\bar{\tau}_{\rm des}&=\frac{\sum_{i=1}^n \tau_0 \cdot \exp \left( \frac{E_\mathrm{des}}{RT_i}\right)A_i}{\sum_{i=1}^n A_i}\\
&=
\begin{cases}
\SI{5.4e6}{\second} & \text{for } \SI{1200}{\joule \per \mole} \\
\SI{1.5e10}{\second} & \text{for } \SI{1400}{\joule \per \mole} \\
\SI{4.1e13}{\second} & \text{for } \SI{1600}{\joule \per \mole} 
\end{cases}.
\end{split}
\end{equation}
The weight $A_i$ is the area of one of the corresponding beamline surface elements with temperature $T_i$ of the COMSOL Multiphysics\textsuperscript{\textregistered} mesh. 
With a fixed desorption energy $E_\mathrm{des}$, the flow reduction can now be calculated on an absolute time scale. 
Since the magnitude of the desorption energy $E_\mathrm{des}$ is not known, a range from $\SI{1200}{\joule\per \mole}$  to $\SI{1600}{\joule\per \mole}$ is investigated. The lower boundary is taken from the estimation given in~\cite{Luo2008}, the upper boundary can be estimated from the first retention measurements which will not be discussed in detail within this publication.  

\subsubsection{Tritium decay \label{sec:betadesorption}}

\begin{figure}[]
	\centering
	\includegraphics[width=\linewidth]{\detokenize{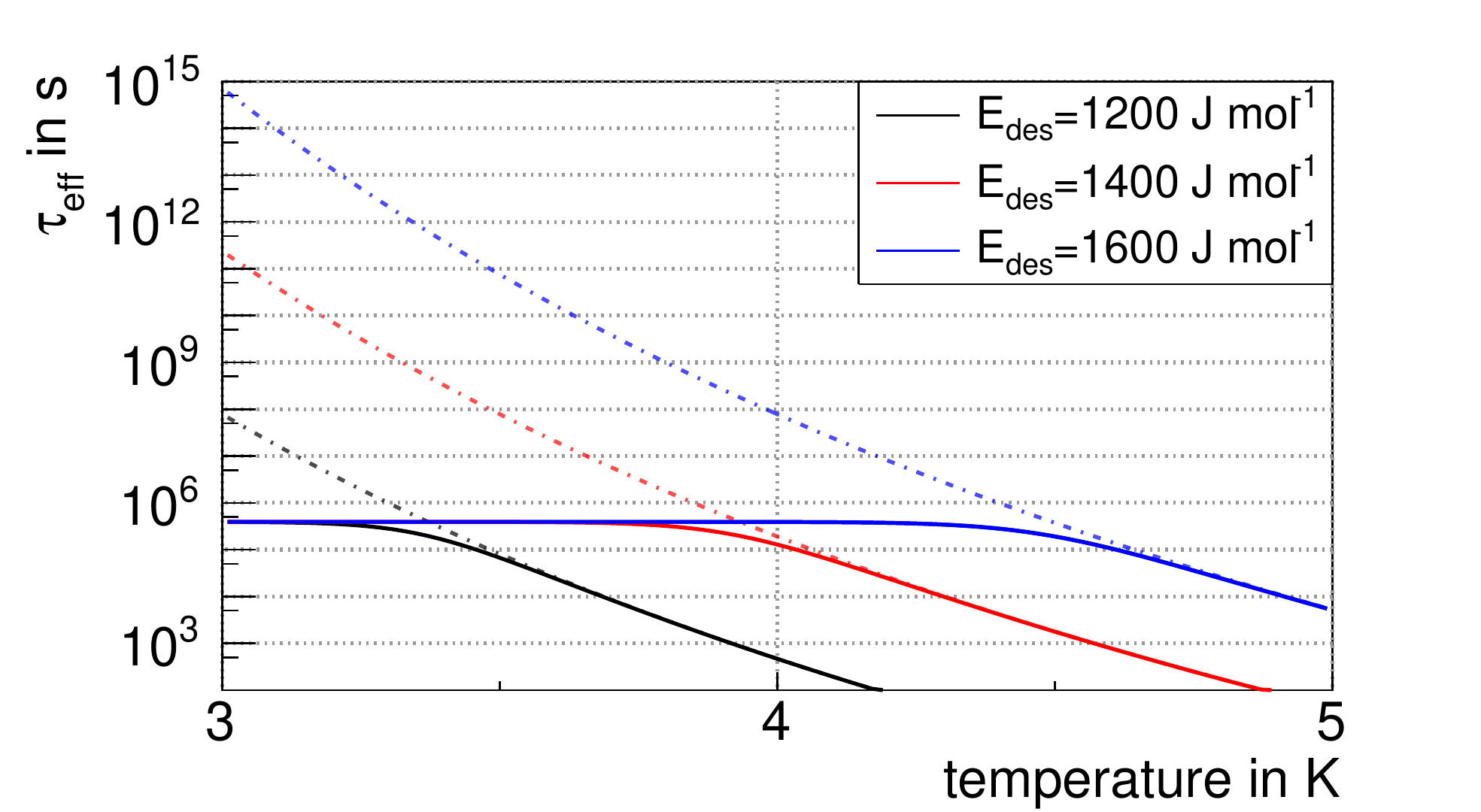}}
	\caption{Course of the effective desorption time for different desorption energies at a surface coverage of $s=\SI{e15}{\TT \per \centi\meter\squared}$ ($\sigma=2$) compared to the case without radioactivity (dashed lines).}
	\label{fig:tau_eff}
\end{figure}

For radioactive adsorbates, the sojourn time can no longer be described solely by the desorption time $\tau_{\mathrm{des}}$ given in Eq.~\ref{eq:tau_des}.
One has to take the influence of radioactive decays inside of the adsorbens into account.
In addition to the released decay products, a tritium decay can also induce the desorption of other atoms in its vicinity, including both tritium and argon. 
The amount of desorbed tritium $\eta(s)$ from a single $\upbeta$-decay inside the argon frost layer depends on the surface coverage $s$. It can be described with the following formula investigated by Malyshev~\cite{Malyshev2008}:

\begin{equation}
\label{eq:eta}
\eta(s)=\eta_{\mathrm{max}}\cdot \frac{s}{s+s_m} \ ,
\end{equation} 
where $\eta_{\mathrm{max}}$ denotes the upper limit for the desorption yield, and $s_m$ the kink between the linear rise and the plateau where $\eta(s)$ reaches saturation.  The values of these parameters are estimated with ${\eta_{\mathrm{max}}=\SI{e3}{T_2\per decay}}$ and
${s_m\approx\SI{4e14}{\TT \per \centi\meter\squared}}$~\cite{Malyshev2008}.
Given this relation, the effective desorption time can be derived from the differential equation

\begin{equation}
\dv{N}{t}=-N\cdot\left( \frac{1}{\tau_{\mathrm{des}}}+\sigma\cdot \eta(s)\cdot \lambda_{\mathrm{T}}\right) \ ,
\end{equation} 
describing the rate of desorbing particles by both thermal desorption and radioactive desorption. The latter is described by the decay constant $\lambda_{\mathrm{T}}$. The variable $\sigma$ is used to describe the gas composition and can take values from 0 to 2 depending on the amount of tritium atoms of the adsorbed isotopologues (H$_2$: $\sigma=0$, HT: $\sigma=1$, T$_2$: $\sigma=2$, ...).
The probability density distribution $\rho(t)$ for the sticking time is then given by

\begin{equation}
\label{eq:p(t)}
\begin{aligned}
\rho(t)=\frac{1}{\tau_{\rm eff}}\cdot\exp\left(-\frac{t}{\tau_{\rm eff}}\right)
\end{aligned}
\end{equation}
with the effective time constant

\begin{equation}
\tau_{\rm eff}=\left(\frac{1}{\tau_{\mathrm{des}}}+\sigma\cdot\eta(s) \cdot \lambda_{\mathrm{T}}\right)^{-1}.
\end{equation}
With the additional decay-induced desorptions, the effective sojourn time will no longer increase with falling temperatures but converges towards a limit as is shown in Fig.~\ref{fig:tau_eff}.

	\section{Semi-Analytical Tracking Model of the CPS \label{sec:analytic_model}}
\label{sec_semianalyticalmodel}
Since tritium decay plays a major role in decreasing the sojourn time of a tritiated molecule inside the cold pump, this effect needs to be taken into account when simulating the tritium reduction factor of the CPS.
The currently available simulation programs for TPMC do not meet the requirements of simulating large reduction factors in combination with radioactive adsorbates and strongly inhomogeneous temperature profiles.
This was the reason for developing a custom C++-based Semi-Analytical Tracking Model. By disassembling the geometry given in Fig.~\ref{fig:cps-model} into its basic geometric primitives (namely cylinders, cones and cuboids), the  motion and interaction points with the surface can be calculated analytically. For the desorption process, a multistage Monte Carlo sampling is needed to determine both the sojourn time and the direction in which a particle desorbs.
The former is done by sampling from the distribution given in Eq.~\ref{eq:p(t)}, the latter with a cosine law sampling~\cite{Greenwood_2002}.
The advantage of the semi-analytical tracking is that the model offers the possibility to integrate three-dimensional models not only for the temperature, but also for the surface coverage along the beamline, which will be discussed in more detail in Sec.~\ref{sec:surfacecoverage}.
The integration of a temperature profile will provide a more precise simulation result than just assuming an average beam tube temperature, since there are regions that are more likely to be hit because of the rotation of the beam tube sections. Therefore a very detailed temperature model from the simulations in Sec.~\ref{sub:beamline_temp} with more than 12\,000 equally distributed temperatures along beam tube sections 2 to 5 of the CPS is used for the simulations.
Compared to the TPMC simulations in Sec.~\ref{ch:CPS-TPMC-model}, the Semi-Analytical Tracking Model requires a new simulation for each desorption energy by changing $\tau_{\rm des}$ and cannot be computed from one set of simulations.

\subsection{Reduction factor calculation}
With the Semi-Analytical Tracking Model, the migration time of a single particle along the entire CPS is calculated.
The simulation of a particle track can have four different outcomes:
\begin{enumerate}[i)]
	\item The particle leaves the CPS through valve V4 into the Pre-Spectrometer. Only these events contribute to the determination of the reduction factor.
	\item The particle is reflected back into the DPS2-F, where it reaches PP4 and is pumped out by the TMP.
	\item The simulation is aborted because the particle takes longer to leave the CPS than the initially set maximal migration time.
	\item The particle decays while still in the CPS, most likely being adsorbed on the argon frost layer.
\end{enumerate}  
Only 1\% of the tritium decays during the nominal operation time of 60 days between subsequent regenerations of the argon frost layer, which is why the loss of tritium due to its decay is neglected in this simulation. The results can also be used for stable isotopes, such as hydrogen or deuterium, if the mean sojourn time $\tau_{\rm des}$ is much longer than the time of flight between two hits of the walls.

To extract the reduction factor of the cold trap, a significant amount of particle tracks and the corresponding migration times need to be calculated. Storing the information into a histogram, normalized to the total number of simulated particle tracks, gives a probability density distribution $m(t)$ for the time $t$ which a molecule needs to migrate along the geometry. Since $m(t)$ depends on the desorption energy, tritium purity, sticking probability and the temperature, these simulations have to be repeated for each parameter setting. 
Once the probability density has been simulated, it can be transformed into a time-dependent tritium flux $\Phi_{\rm out}(t)$ from the downstream end of the CPS to the Pre-Spectrometer. This is done by integrating over the migration probability $m(t)$ and multiplying with the expected constant tritium flux $\Phi_{\rm in}=\SI{e12}{\molecules\per \second}$~\cite{PhDJansen2015} from the DPS2-F into the CPS
	\begin{equation}
	\label{eq:flow}
	\Phi_{\rm out}(t)=\Phi_{\rm in}\int_{0}^{t} m(t')\ \rm dt'\ .
\end{equation}
The reduction factor $R(t)$ after time $t$ of continuous operation is defined as the ratio of the outgoing and incoming flux:
	\begin{equation}
\label{eq:reduction}
R(t)=\frac{\Phi_{\rm in}}{\Phi_{\rm out}(t)}=\left(\int_{0}^{t} m(t')\ \rm dt'\right)^{-1}\ .
\end{equation}
\begin{figure}[]
	\centering
	\begin{subfigure}[c]{0.41\textwidth}
		
		\includegraphics[width=\textwidth]{\detokenize{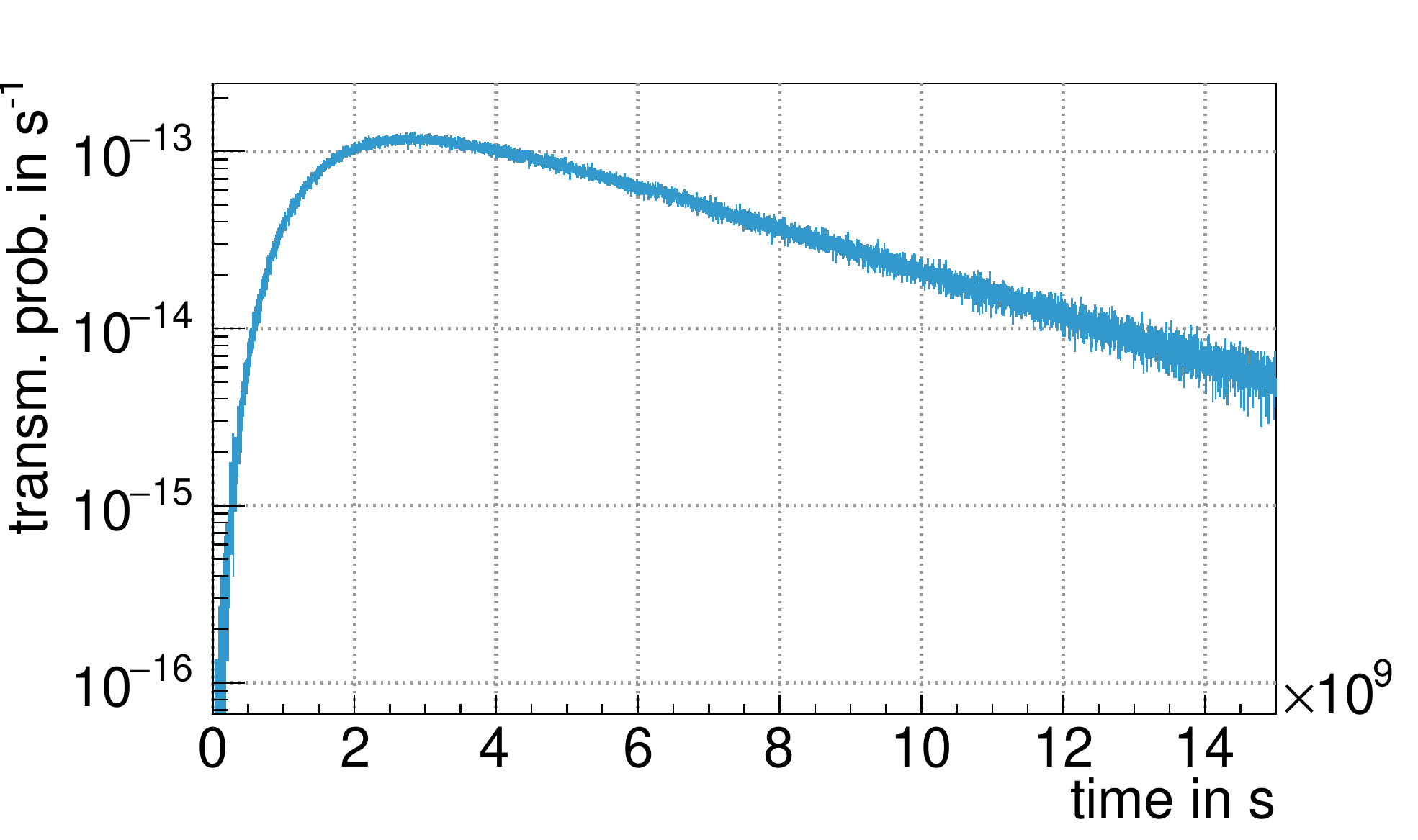}}
		\subcaption{}
		
	\end{subfigure}
	\begin{subfigure}[c]{0.41\textwidth}
		\includegraphics[width=\textwidth]{\detokenize{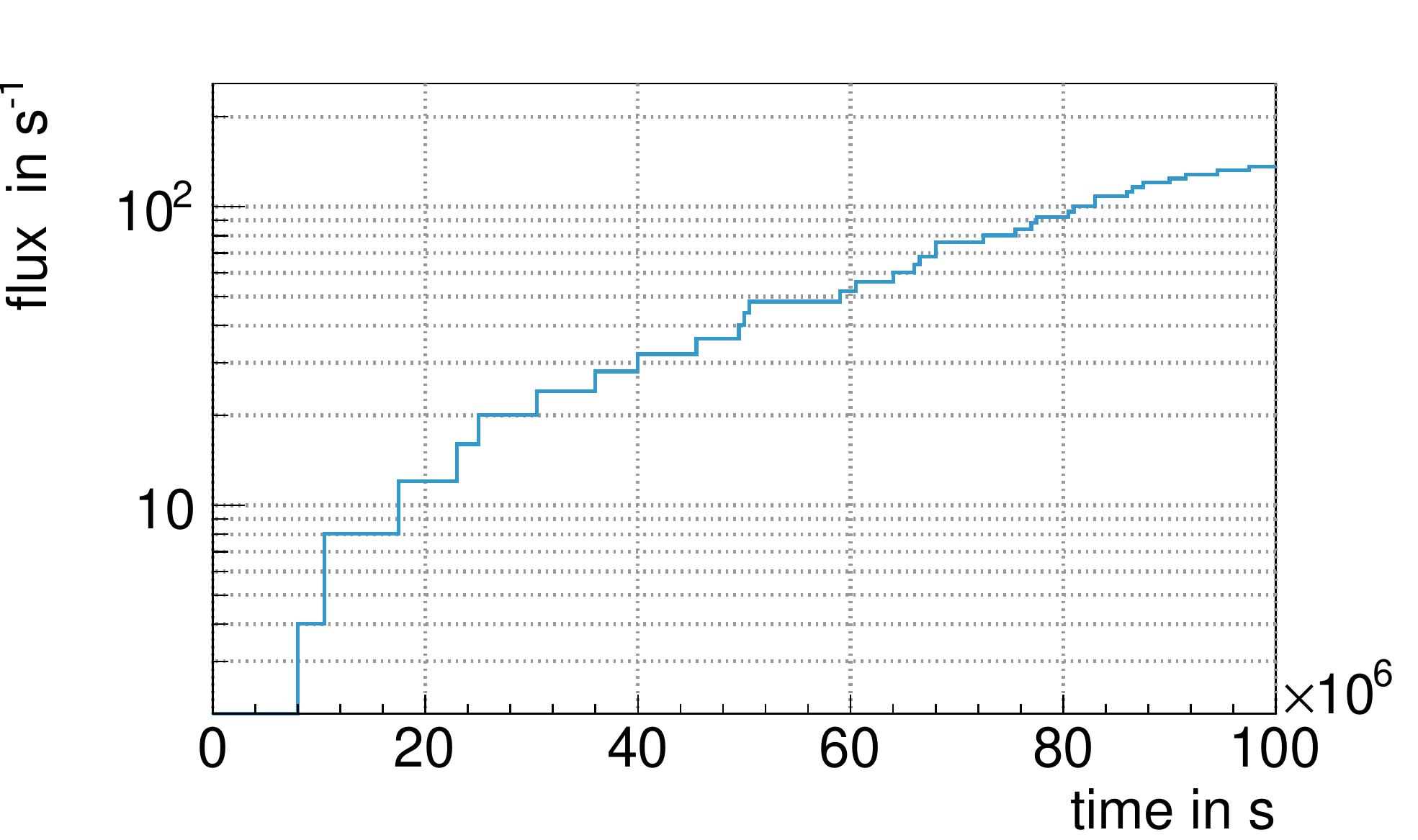}}
		\subcaption{}
		
	\end{subfigure}
	\caption{ (a) Transmission probability function for a desorption energy of $\SI{1200}{\joule\per \mole}$ with pure tritium ($\sigma=2$); (b) zoom to the region of interest of the resulting tritium flux into the Pre-Spectrometer according to Eq.~\ref{eq:flow}.}
	\label{fig:m(t)}
\end{figure}

The transmission probability density and its integral (Eq.~\ref{eq:flow}) are shown in Fig.~\ref{fig:m(t)}. Although the distributions were simulated with up to $\SI{2.5e11}{}$ events each, the simulations  provide virtually no events in the region between 0 and 60 days for reduction factors ${R(60\,{\rm d}) > 10^{12}}$. In this case a linear interpolation between the first non zero bin and the origin (where the tritium flow is expected to be zero) is applied.

	\section{Results \label{sec:results}}
In the following, the simulation results of the flow reduction of both DPS2-F and CPS are presented.
In Secs.~\ref{sec:TPMC_models} and~\ref{sec_semianalyticalmodel}, two different approaches were introduced. 
The DPS2-F results are solely based on MolFlow+ simulations, while the CPS results were complemented by the Semi-Analytical Tracking Model.
\subsection{DPS2-F}
The DPS2-F simulation has been split into four parts as described in Sec.~\ref{sec:dpsmodel}. Table~\ref{tab:dpsresults} gives an overview of the results. 
\begin{table}
\centering
\caption{Results of the four parts of the DPS2-F gas flow simulation with MolFlow+.}
\resizebox{\linewidth}{!}{
\begin{tabular}{cccccc}
\hline
Part & Desorption & Inlet \& outlet & Inlet counts & Outlet counts \\ 
\hline
1 & $E_0$ & $E_0$ \& $G_2$ & 36901582 & 466310 \\ 
2 & $G_1$ & $G_2$ \& $G_3$ & 3385527 & 342948 \\ 
3 & $G_2$ & $G_3$ \& $G_4$ & 809346 & 82173 \\ 
4 & $G_3$ & $G_4$ \& $V_2$ & 2142160 & 104502 \\ 
\hline
\end{tabular}
}
\label{tab:dpsresults}
\end{table} 
Concatenating all four simulations to an overall reduction factor yields
\begin{equation}
R_{\rm DPS2-F}=(1.577 \pm 0.008_{\rm stat.}) \times 10^5
\end{equation} 
with the statistical uncertainty calculated by using binomial statistics. 
To estimate the systematic uncertainty of the TMP pumping probability (see Sec.~\ref{sec:dpsmodel}) a dedicated simulation with $\alpha=0.202$ was performed.
The result
\begin{equation}
	R_{\rm DPS2-F}^{\mathrm{lower}}=(8.99 \pm 0.05_{\rm stat.}) \times 10^4
\end{equation}
gives a lower limit for the reduction factor with a 20\% reduction of pumping probability.

\subsection{CPS}
\subsubsection{MolFlow+}
Two different scenarios were simulated with MolFlow+. The first one is the standard neutrino mass measurement; the parameter of interest is the reduction factor. The other scenario is the commissioning measurement with D$_2$; the parameters of interest are the pressures at both pump ports. 
In the commissioning simulation, the inlet valve V2 and the outlet valve V4 are assumed to be closed, while they are opened during the neutrino mass measurement simulation. 

Combining the resulting values $R(t)$ for the neutrino mass measurement simulation with $p_{\text{PP1}}(t)$ and $p_{\text{PP2}}(t)$ for the commissioning measurement simulation, the factor $k(t)$ of Eq.~\ref{eq:reduction-pressure} was determined. 

\begin{figure}[]
\centering
\includegraphics[width=\linewidth]{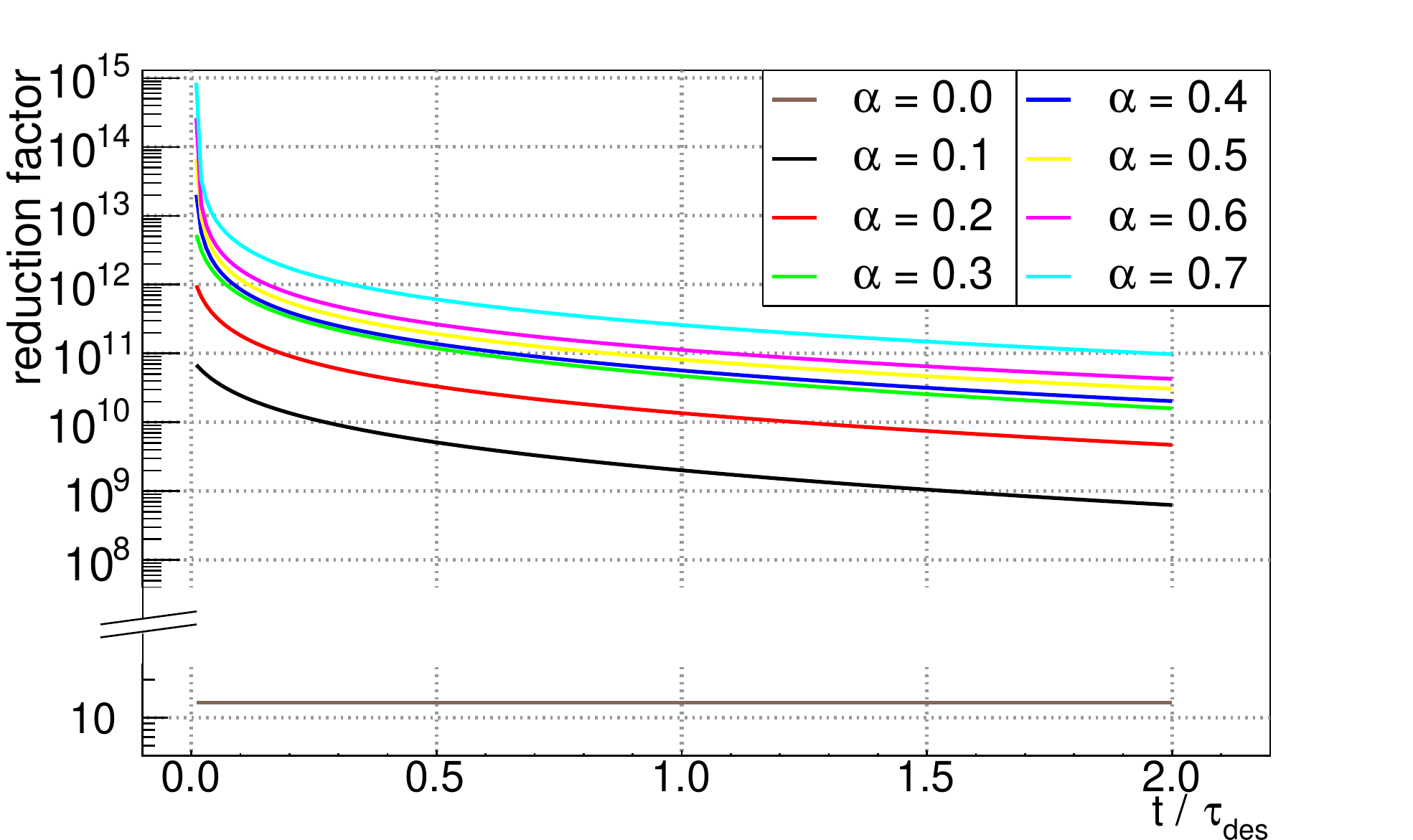}
\caption{Reduction factors for standard tritium operation and for different sticking coefficients of the argon frost layer from ${\alpha_{\text{Ar}}=0.0 \text{ to } 0.7}$, simulated with MolFlow+. For  $\alpha_{\text{Ar}}= \nobreak 0.0$, the value is constant at $R \approx 13.2$.}
\label{fig:R-alpha01234567}
\end{figure}

\begin{figure}[]
\centering
\includegraphics[width=\linewidth]{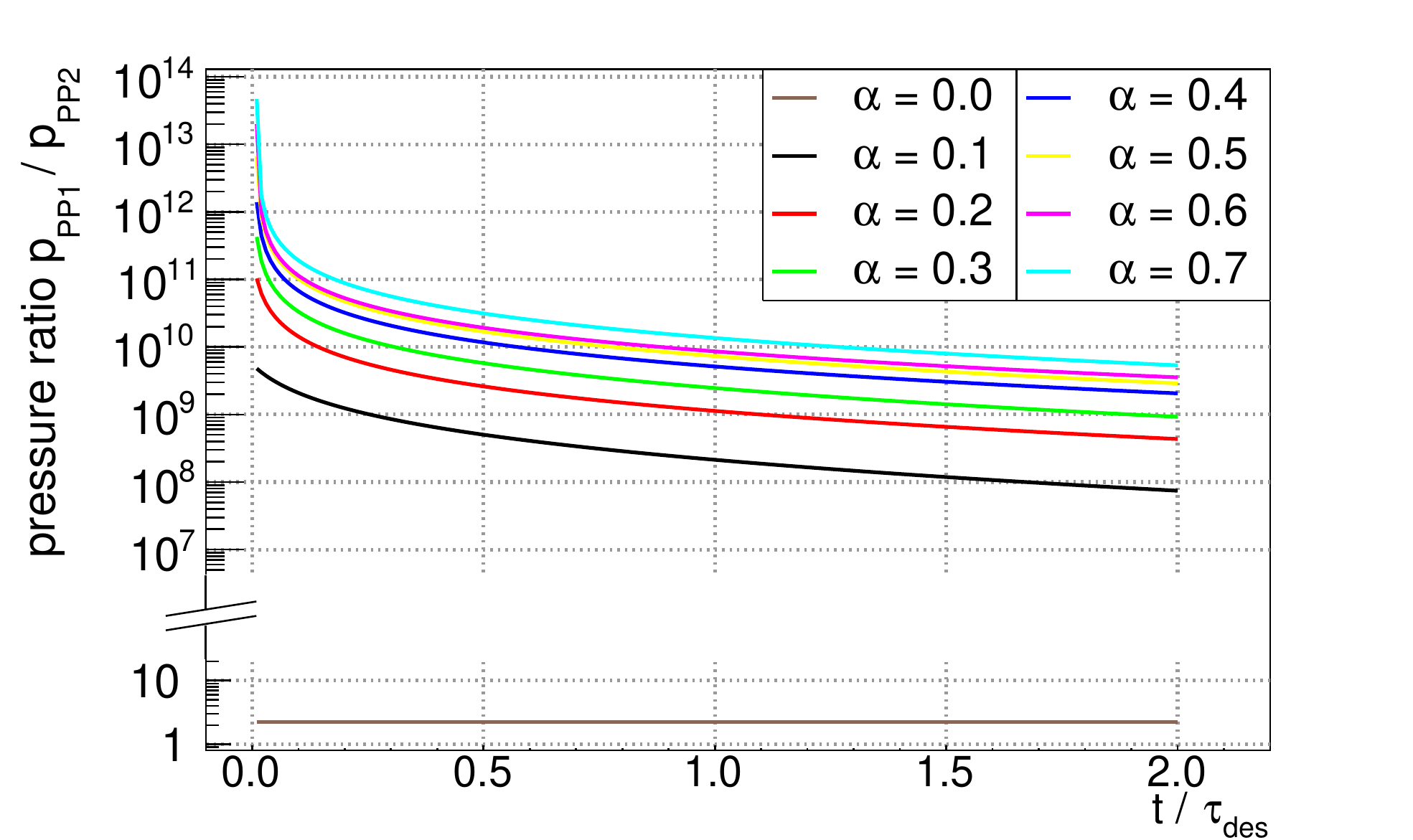}
\caption{Pressure ratios of pump port 1 to pump port 2 for the D$_2$ commissioning measurement scenario and for different sticking coefficients of the argon frost layer from $\alpha_{\text{Ar}}=0.0 \text{ to } 0.7$, simulated with MolFlow+. For  $\alpha_{\text{Ar}}=0.0$, the value is constant at $p_{\text{PP1}}/p_{\text{PP2}} \approx 2.2$.}
\label{fig:p-alpha01234567}
\end{figure}

\begin{figure}[]
\centering
\includegraphics[width=\linewidth]{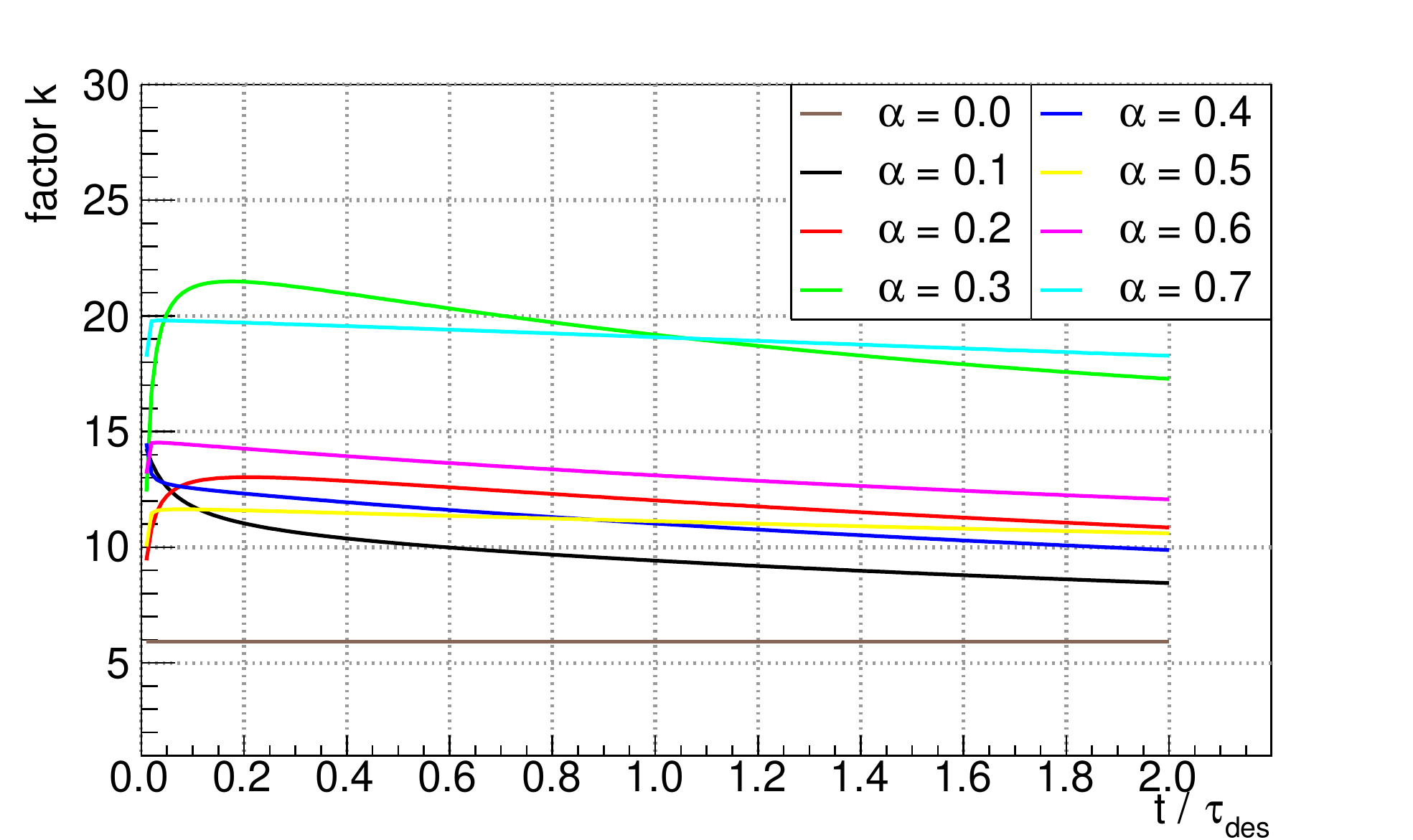}
\caption{Factor $k=R/(p_{\text{PP1}}/p_{\text{PP2}})$ for different sticking coefficients from $\alpha_{\text{Ar}}=0.0 \text{ to } 0.7$, simulated with MolFlow+. Here $R$ is the reduction factor for standard tritium operation and $p_{\text{PP1}}/p_{\text{PP2}}$ is the pressure ratio for the  D$_2$ commissioning measurement scenario. For  $\alpha_{\text{Ar}}=0.0$, the value is constant at $k \approx 5.9$.}
\label{fig:k-alpha01234567}
\end{figure}

The simulation results for the three relevant parameters are displayed in Figs.~\ref{fig:R-alpha01234567} to~\ref{fig:k-alpha01234567}. 
Since the mean sojourn time $\tau_{\rm des}$ is unknown, the time axes are normalized to $\tau_{\rm des}$. The time interval for the iterative integration was set to $\Delta t=0.01 \cdot \tau_{\rm des}$. 
As expected, the reduction factor and the pressure ratio show a similar time-dependent behavior. Over a period of $2\tau_{\rm des}$  both parameters decrease by 2 to 3 orders of magnitude, except for $\alpha=0.0$ where they stay constant as there is no cryosorption at all. 
Lower sticking coefficients result in lower reduction factors and pressure ratios. The results for the ad hoc factor $k(t)$ are important for interpreting the D$_2$ commissioning measurements. The simulated values lie between 8.5 and 21.5, and stay more or less constant. 

\begin{figure}[]
	\centering
	\includegraphics[width=\linewidth]{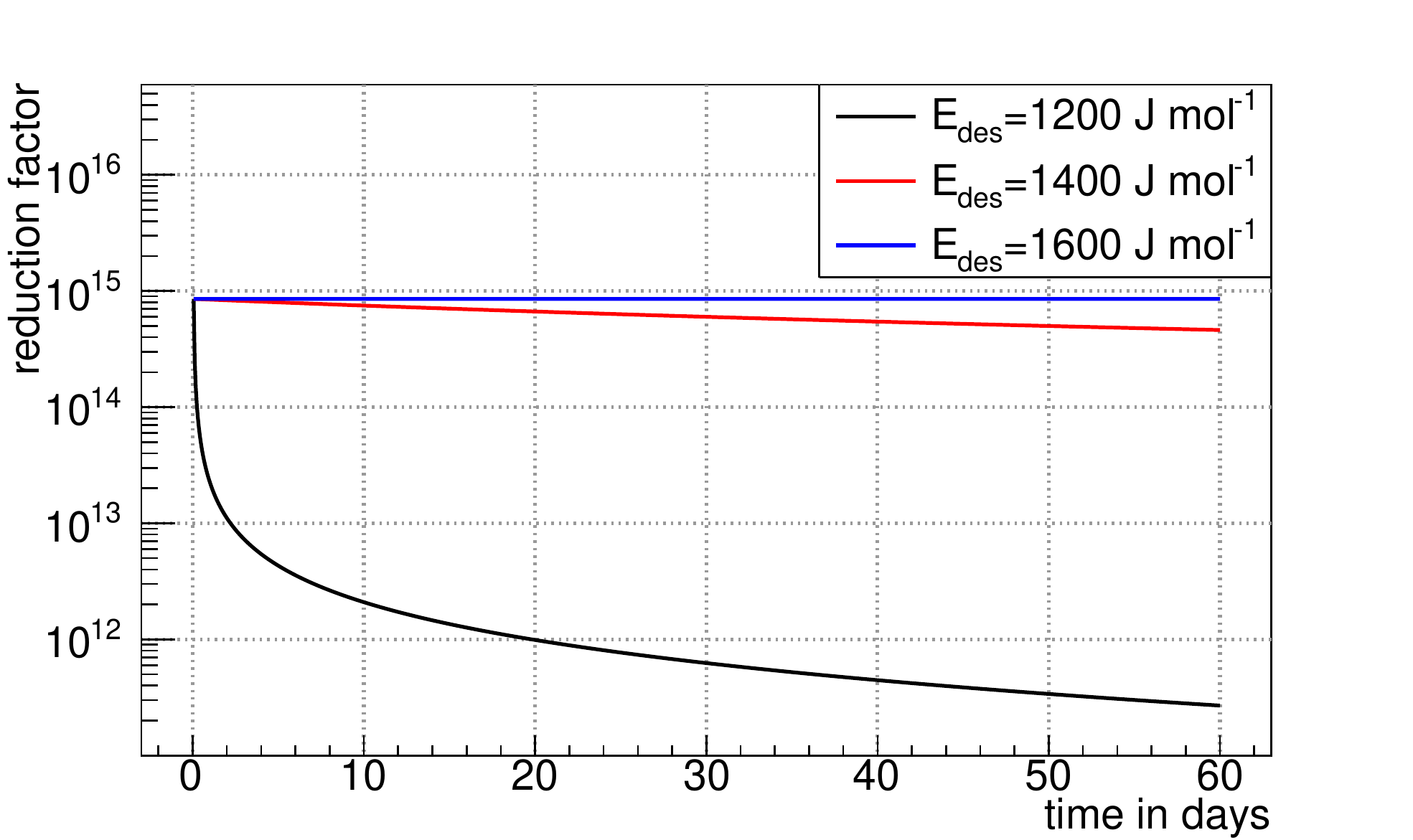}
	\caption{The reduction factor for three different desorption energies $E_\mathrm{des}$ and $\alpha = 0.7$ with the inhomogeneous CPS beam tube temperature profile, simulated with MolFlow+.
	The effect of tritium migration can be clearly seen for the lower desorption energies.}
	\label{fig:R-1200-1400-1600}
\end{figure}

For the nominal KATRIN operation sufficient tritium suppression for the whole 60-day run period is of paramount importance. To understand the long-term suppression, the x-axis in Fig.~\ref{fig:R-alpha01234567} has to be multiplied with a constant $\tau_{\mathrm{des}}$. Therefore, the desorption energy $E_\mathrm{des}$ is fixed, and the inhomogeneous beamline temperature profile is included according to Sec.~\ref{sub:includetempprofile}. This has been done for $\alpha=0.7$ and three different desorption energies ${E_\mathrm{des}=\SI{1200}{\joule\per \mole},}\SI{1400}{\joule\per \mole}\text{, and } \SI{1600}{\joule\per \mole}$. The results are shown in Fig.~\ref{fig:R-1200-1400-1600}.

\subsubsection{Semi-Analytical Tracking Model}
\label{sec:surfacecoverage}

\begin{figure*}[]
	\centering
	\includegraphics[width=\linewidth]{\detokenize{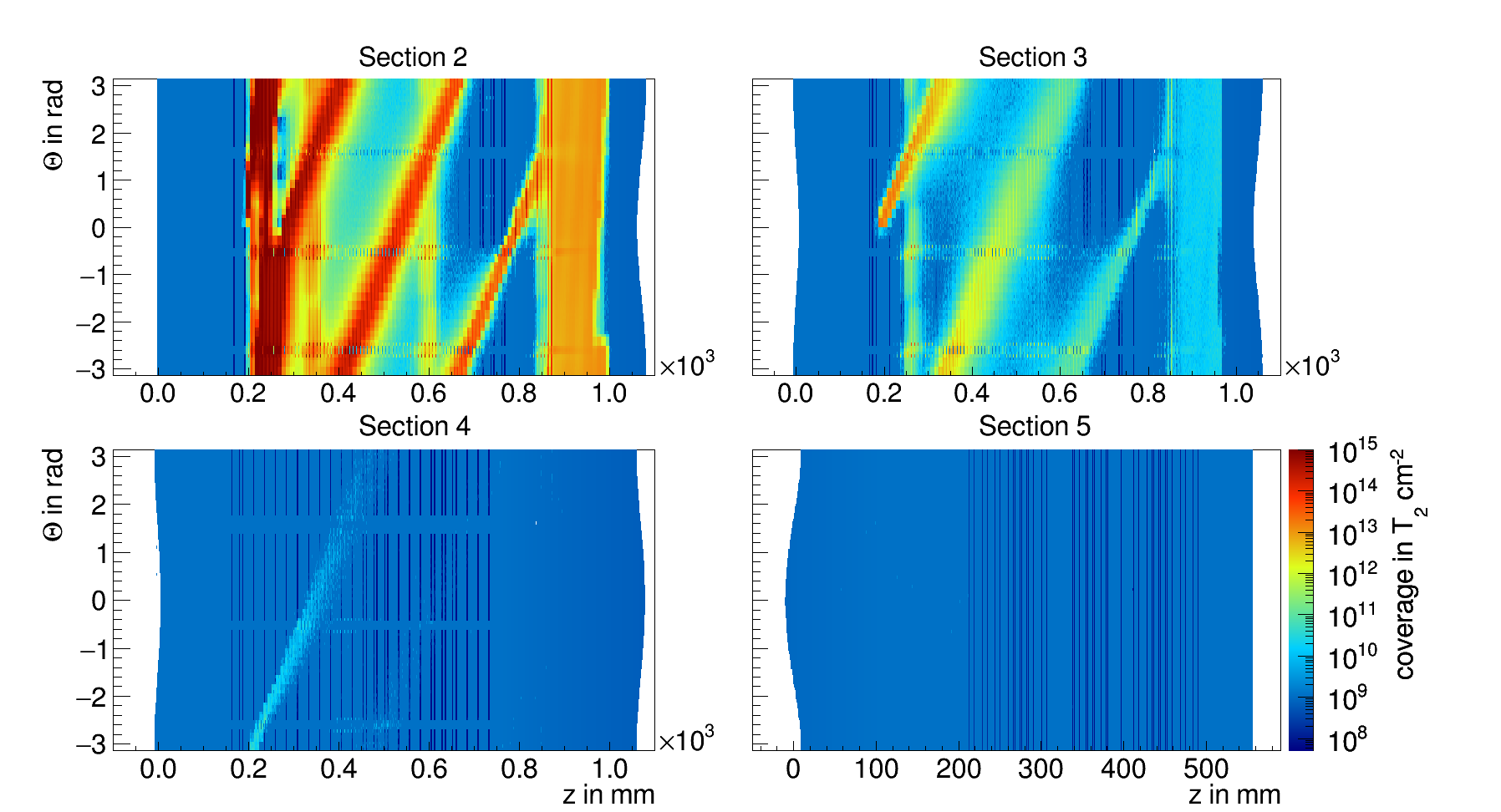}}
	\caption{Two-dimensional representation of the tritium coverage after a pumping time of 60 days, simulated with the Semi-Analytical Tracking Model. The z axis follows the central axis of the beam tube. The azimuthal angle $\Theta$ covers the full circumference of the beam tube. The apparent inhomogeneities are due to the inhomogeneous temperature profile along the cryogenic pumping section.}
	\label{fig:distrBT2-5}
\end{figure*}

\begin{figure}[]
	\centering
	\includegraphics[width=\linewidth]{\detokenize{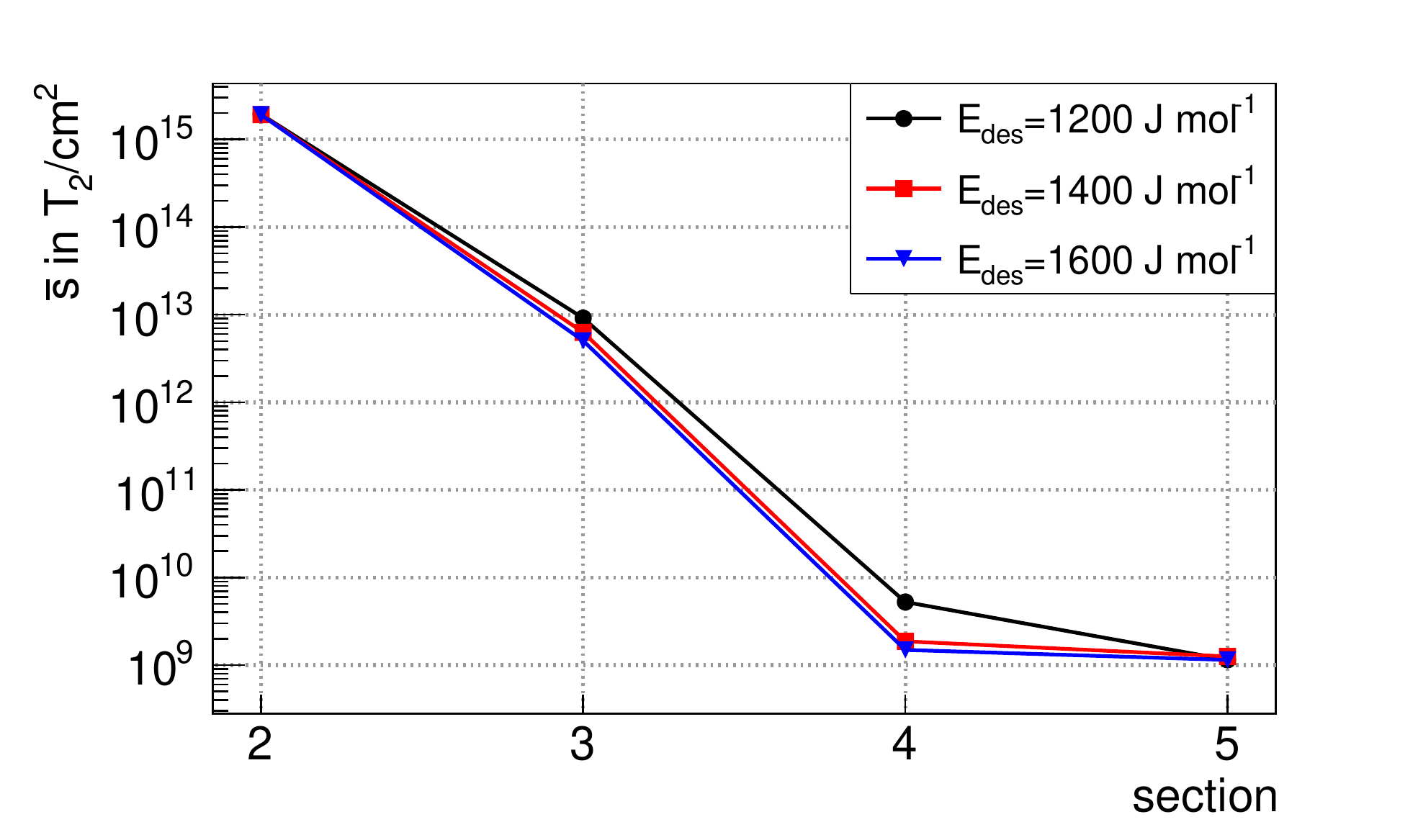}}
	\caption{Distribution of the mean surface coverage $\bar{s}$ per beam tube along the CPS, simulated with the Semi-Analytical Tracking Model. It shows an almost exponential decrease. The values for segment 5 have to be seen as an upper limit. Despite the huge number of $\SI{2.5e11}{}$ simulated molecules for each setting, no significant amount of molecules reached this section.}
	\label{fig:meanSurfCov}
\end{figure}

Because of the dependence of $\tau_\mathrm{eff}$ on $\eta(s)$, a detailed knowledge of the surface coverage $s$ along the segments of the CPS is required to simulate the impact of radioactive decays on the tritium reduction factor.
With the cold trap temperature profile implemented in the simulation code, the surface coverage turns from a smooth distribution into the density map shown in Fig.~\ref{fig:distrBT2-5}. The correlation between the lower temperatures and a high surface coverage can be easily explained by the mean sojourn time of the molecules adsorbed on the argon frost, which depends strongly on the local temperature (see Eq.~\ref{eq:p(t)}). The inhomogeneous temperature profile leads to an enhanced migration from regions of higher temperatures to regions of lower temperatures.
As shown in Fig.~\ref{fig:meanSurfCov}, the mean surface coverage 
\begin{equation}
\bar{s}=\frac{\sum_i s_i\cdot n_i}{n_{\mathrm{tot}}}
\end{equation} decreases almost exponentially from about $\SI{e15}{\TT \per \centi\meter\squared}$ at the upstream entrance  to  $\SI{e9}{\TT \per \centi\meter\squared}$ or less at the downstream end of the CPS.
Here $s_i$ denotes the surface coverage of bin $i$ in Fig.~\ref{fig:distrBT2-5} and $n_i$ the corresponding amount of molecules. Calculating the average weighted by particles within a given bin instead of its area is required to correctly simulate $\upbeta$-desorptions as described in Sec.~\ref{sec:betadesorption}.

To cover the expected range of the desorption energy (see Sec.~\ref{sub:includetempprofile}), a set of simulations reaching from $E_{\mathrm{des}}=\SI{1200}{\joule\per \mole}$  to $E_{\mathrm{des}}=\SI{1600}{\joule\per \mole}$ has been performed.
For non-radioactive gases with desorption energies above $\SI{1200}{\joule\per \mole}$, this simulation method is not suitable since the probability function used to calculate the reduction (see Eq.~\ref{eq:reduction}) has no entries close to the region from $t=\SI{0}{days}$ to $t=\SI{60}{days}$. Therefore the uncertainty of any extrapolation would span several orders of magnitude. 

\begin{figure}[]
	\centering
	\includegraphics[width=\linewidth]{\detokenize{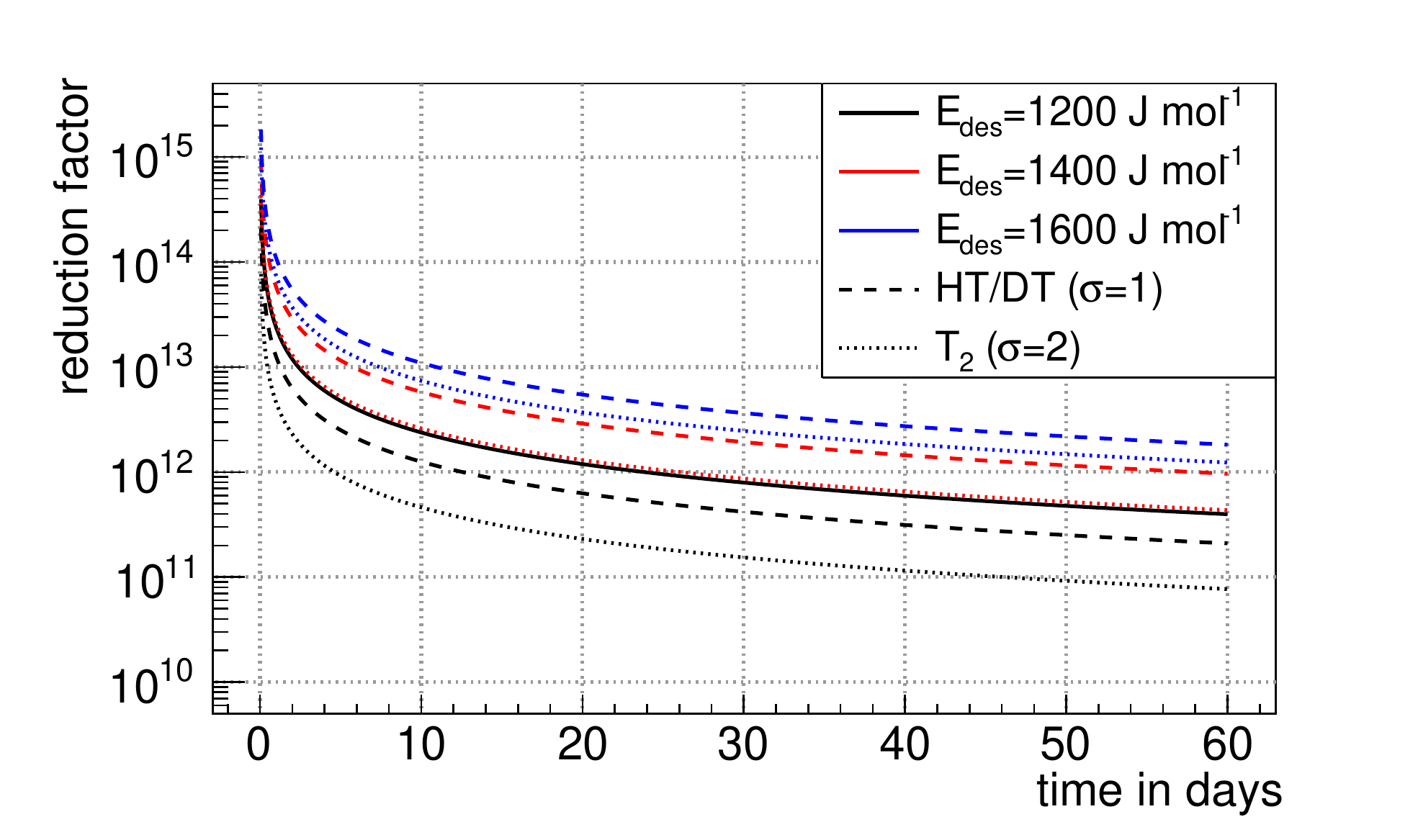}}
	\caption{Time-dependent tritium reduction when assuming no radioactivity for a desorption energy of $\SI{1200}{\joule\per \mole}$ (solid line) and for the case of a tritiated adsorbate for even higher desorption energies (dashed/dotted lines), simulated with the Semi-Analytical Tracking Model. The composition of the adsorbate is described with the variable $\sigma$ which is 1 for $\mathrm{HT}$/$\mathrm{DT}$ and 2 for pure $\mathrm{T_2}$. }
	\label{fig:retention_plot}
\end{figure}

But as soon as radioactive desorption is considered, the sojourn time and the reduction factor are drastically reduced. In this case, the Semi-Analytical Tracking Model is valid even for higher desorption energies and lower temperatures where $\tau_{\rm eff}$ converges towards a constant value, as shown in Fig.~\ref{fig:tau_eff}.
This can be seen in Fig.~\ref{fig:retention_plot} where the tritium reduction factor rises slower with larger desorption energies for T$_2$ than it does for isotopologues with only one tritium atom (HT, DT).

In order to reduce the complexity of the simulation code, a static mean surface coverage $\bar{s}$ after 60 days of each beam tube section is used. These values were obtained from preceding simulations for each binding energy without taking radioactive desorptions into account (see Fig.~\ref{fig:meanSurfCov}).
Using $\bar{s}$ is justified, since the influence of radioactive decays is only significant in regions with low temperature (see Fig.~\ref{fig:tau_eff}), where the majority of the molecules is adsorbed. Because no time dependency is implemented, the results for the reduction factor have to be seen as a conservative lower limit.
Here the influence of the desorption energy on the reduction factor is only in the range of two orders of magnitude. Even for the very conservative assumption of a static surface coverage and a desorption energy of $\SI{1200}{\joule\per \mole}$, the simulation yields a reduction factor of at least $\SI{2.6e+10}{}$, which exceeds the requirements by three orders of magnitude.
\begin{table}
	\centering
	\caption{Overview of the simulation results of the reduction factors for different isotopologues after 60 days, simulated with the Semi-Analytical Tracking Model. The reduction factors marked with a star are extrapolated values with a lower limit of $\SI{2.5e11}{}$.}
	\resizebox{\linewidth}{!}{
\begin{tabular}{cccc}
	\hline
	$E_\mathrm{des}$ in $\si[per-mode=reciprocal]{\joule \per \mole}$& H$_2/$D$_2$& HT$/$DT & T$_2$ \\ 
	\hline 
$ 1200$& $\SI{4.0e+11}{}^{\ast}$ & $\SI{2.1e+11}{}$ & $\SI{7.7e+10}{}$ \\
$ 1400$& -- & $\SI{9.6e+11}{}^{\ast}$ & $\SI{4.3e+11}{}^{\ast}$ \\
$ 1600$& -- & $\SI{1.8e+12}{}^{\ast}$ & $\SI{1.2e+12}{}^{\ast}$ \\

	\hline 
\end{tabular}
} 
\label{tab:suppression}
\end{table}

	\section{Discussion}
\label{sec:discussion}

With the specified reduction factor of $10^5$ for the DPS2-F, the results obtained 
with the MolFlow+ simulations are right on target. The lower limit for the reduction 
factor takes into account the conservative uncertainty of the TMP pumping probabilities for T$_2$.

On the other hand, the MolFlow+ results for the CPS show a reduction factor that 
exceeds the specified goal of $10^7$ by several orders of magnitude.
For the safe operation of KATRIN, it is important to understand the accuracy of these results.

The impact of the cold trap capacity on the sticking probability has not been included in 
the model. After 60 days of KATRIN operation, approximately 1.7\% of the total cold trap capacity is reached~\cite{PhDJansen2015}. 
Further, it is assumed that the first 1.7\% of the trap is covered to 100\% ($\alpha = 0$), while 
the rest is free of adsorbates. Under the assumption, the reduction factor drops exponentially along the cold 
trap; the final value of $2.7 \times 10^{11}$ (see Fig.~\ref{fig:R-1200-1400-1600}) for the lower 
limit of $E_{\mathrm{des}}$ would be reduced by 36\%.

A second source of systematic uncertainties is the concatenation algorithm of the four 
independently simulated parts of the CPS MolFlow+ geometry. By comparing simulations with a small 
sticking coefficient $\alpha = 0.1$ for a single-pass simulation and with concatenation, the total 
error is estimated to be less than a factor of two.
 
The analysis of the time dependence with MolFlow+ simulations provides results 
for a time scale normalized to the sojourn time $\tau_{\rm des}$. In order to extract results on 
an absolute time scale, an appropriate range for the unknown desorption energy $E_\mathrm{des}$ has 
to be estimated. Reasonable values lie between $\SI{1200}{\joule \per \mole}$ and 
$\SI{1600}{\joule \per \mole}$. In addition the temperature inhomogeneity of the cold trap was 
included by calculating an effective mean sojourn time $\bar{\tau}_{\rm des}$ 
(see Sec.~\ref{sub:includetempprofile}). This is justified by the large number of adsorptions 
on different segments when particles migrate through the cold trap.

Considering all of the relevant systematic uncertainties, the results' accuracy is assumed to be on the 
order-of-magnitude level. The simulations are very important for characterization measurements of 
the CPS cold trap with D$_2$ because the reduction factor cannot be measured directly. The 
measurements might also help to reduce the large uncertainties of the input parameters. For the 
standard KATRIN operation, additional $\upbeta$-induced desorptions from tritium decays have to 
be taken into account, which would reduce the assumed sojourn time even further. Despite all these 
uncertainties in the MolFlow+ simulations, the expected reduction factor still exceeds the nominal 
value by several orders of magnitude. 

Compared to the D$_2$ simulations with MolFlow+, the Semi-Analytical Tracking Model for 
tritiated isotopologues has a significantly lower reduction factor, as shown in 
Fig.~\ref{fig:retention_plot}. The reduction factor for tritium still exceeds the requirements by far.
A necessary simplification had to be made in order to attain the reduction factor for radioactive adsorbates. This simplification includes the assumption of a time independent, mean surface coverage per beam tube section, as described in Sec.~\ref{sec:surfacecoverage}.
This approach overestimates the surface coverage and results in a conservative limit.

Another systematic uncertainty arises from the linear interpolation of the cumulative  probability density function of the migration time (see Fig.~\ref{fig:m(t)}) for $t<\SI{60}{days}$.
If no event is produced in this region, a conservative lower limit of $\SI{2.5e11}{}$ for the reduction factor can be inferred. 

\begin{figure}
	\centering
	\includegraphics[width=\linewidth]{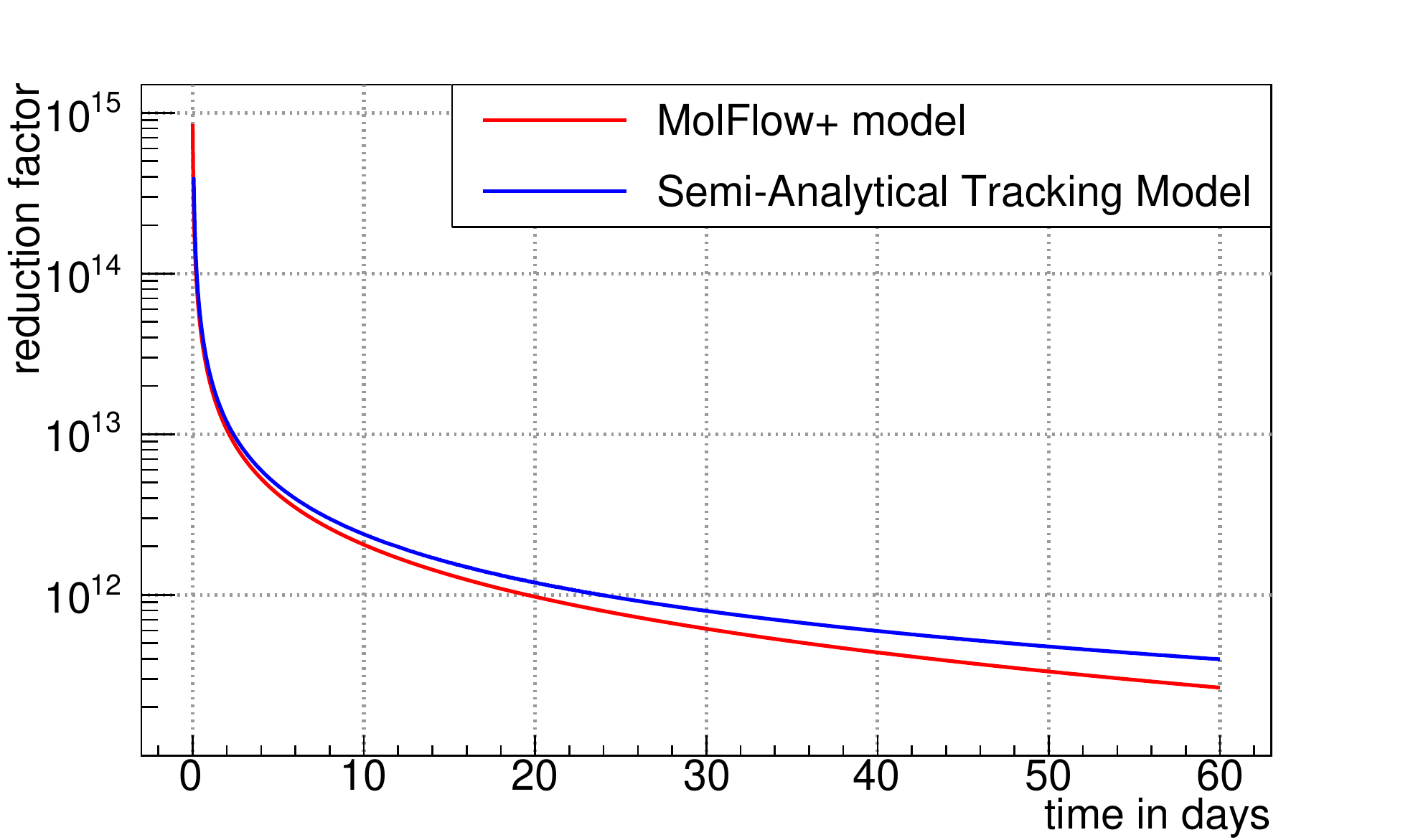}
	\caption{CPS reduction factor results on an absolute time scale for $\alpha=0.7$ and $E_{\mathrm{des}}=\SI{1200}{\joule \per \mole}$. Both the MolFlow+ model and the Semi-Analytical Tracking Model are shown for comparison. }
	\label{fig:comparison_lutz_fabian}
\end{figure}

In Fig.~\ref{fig:comparison_lutz_fabian}, the results of both simulation methods are compared for 
$D_2$ simulations for an assumed desorption energy of $\SI{1200}{\joule \per \mole}$. The two very different approaches show similar results. After 60 days, the results $R=\SI{2.7e11}{} $ for 
MolFlow+ and $R=\SI{4.0e11}{}$ for the Semi-Analytical Tracking Model agree to within a factor of two. Taking into account the complexity and different approximations of both methods, the results can still be considered to be in good agreement. 

	\section{Conclusions\label{sec:conclusions}}
In the KATRIN experiment, it is mandatory to reduce the tritium gas flow between the WGTS and 
the Pre-Spectrometer by at least 14 orders of magnitude. 
Tritium decays inside the spectrometer section would otherwise increase the background 
rate of the experiment, limiting the ultimate sensitivity for the neutrino mass.
For this reason, the differential pumping section DPS2-F and the cryogenic pumping section 
CPS are located between the WGTS and the spectrometer section to reduce the tritium flow accordingly.

An initial simulation of the temperature profile of the CPS cold trap with COMSOL Multiphysics\textsuperscript{\textregistered} revealed 
inhomogeneities that were taken into account in the subsequent vacuum simulations.
For these simulations in the molecular flow regime, two different models were used: a TPMC simulation with MolFlow+ and a Semi-Analytical Tracking Model developed in C++.

The simulations help to infer actual flow reduction factors from the measured pressure ratios close to the inlet and outlet flanges of the pumping sections. The initial simulations of the cold trap of the CPS with deuterium included time-dependent migration along the beamline due to thermal re-desorption. The effect of $\upbeta$-decay-induced desorption was added by decreasing the desorption time $\tau_\mathrm{des}$ accordingly. Even with the large systematic uncertainties due to the wide spread of the possible input parameters in the simulations, both methods reach the same conclusion that the combined reduction factor of the pumping sections exceeds the design value by several orders of magnitude.   

The preliminary results of the ongoing measurements with deuterium support the findings of these simulations. Further measurements and detailed comparisons with the models will help us to reduce the uncertainties of the input parameters, which would ultimately lead to more accurate predictions.
	\section{Acknowledgements}
We acknowledge the support of the Helmholtz Association (HGF), 
the German Ministry for Education and Research BMBF (05A17VK2), 
the Helmholtz Alliance for Astroparticle Physics (HAP), 
the Helmholtz Young Investigator Group VH-NG-1055, 
the Research Training Group (GRK1694),
and the DFG graduate school KSETA (GSC~1085).

	\bibliographystyle{elsarticle-num}
	\bibliography{bibliography}
	
\end{document}